\documentclass[prd,preprint,superscriptaddress,amsmath,amssymb,nofootinbib]{revtex4}
\usepackage{graphicx}% Include figure files
\usepackage{dcolumn}% Align table columns on decimal point
\usepackage{bm}% bold math
\usepackage{amssymb}
\usepackage{amsmath}
\usepackage{epsfig}    
\usepackage{color}
\usepackage{slashed}
\usepackage{hhline}
\usepackage{multirow}
\def\be{\begin{equation}}
\def\ee{\end{equation}}
\newcommand{\bea}{\begin{eqnarray}}
\newcommand{\eea}{\end{eqnarray}}
\begin{document}

 \begin{flushright} {KIAS-P19042}  \end{flushright}

%%%%%%%%%
\title{Lepton specific two Higgs doublet model based on $U(1)_X$ gauge symmetry with dark matter}

\author{Takaaki Nomura}
\email{nomura@kias.re.kr}
\affiliation{School of Physics, KIAS, Seoul 02455, Republic of Korea}

\author{Prasenjit Sanyal}
\email{psanyal@iitk.ac.in}
\affiliation{Department of Physics, Indian Institute of Technology Kanpur, Kanpur 208016, India}

\date{\today}

\begin{abstract}
We discuss a two Higgs doublet model with extra $U(1)_X$ gauge symmetry where lepton specific (type-X) structure for Yukawa interactions is realized by charge assignment of fields under the $U(1)_X$. Extra charged leptons are introduced to cancel gauge anomaly associated with extra gauge symmetry. In addition, we introduce scalar fields as dark matter candidates to which we assign $Z_2$ odd parity for guaranteeing stability of them. We then analyze phenomenology of the model such as scalar potential, muon anomalous magnetic dipole moment, collider physics associated with $Z'$ boson from $U(1)_X$, and dark matter physics.
Carrying out numerical analysis we search for phenomenologically viable parameter region.   
 \end{abstract}
\maketitle

\section{Introductions}

The standard model (SM) of particle physics has been very successful to explain experimental results and its particle contents are confirmed 
completely by the discovery of the Higgs boson at the Large Hadron Collider (LHC).
Although the SM is quite successful, there can be a new physics beyond the SM (BSM) accommodating with the experimental data 
and it is motivated by several issues such as existence of dark matter(DM) and non-zero mass of neutrinos which can not be explained within the SM.
Furthermore the existence of new physics would induce interesting phenomenology such as flavor physics and new particle signatures at collider experiments.
 
One of the interesting extension of the SM is two Higgs doublet model (THDM) in which a second Higgs doublet is introduced.
In general, THDM has flavor changing interactions through Yukawa interactions of both quarks and leptons, which are strongly constrained 
by various experiments searching for flavor violating processes.
In many approaches softly broken $Z_2$ symmetry is introduced to restrict Yukawa interactions to avoid flavor changing neutral current(FCNC).
One can also apply an extra $U(1)$ gauge symmetry to control Yukawa interactions associated with two Higgs doublets. 
In such a scenario rich phenomenology would be induced from scalar bosons from Higgs sector as well as $Z'$ boson from extra $U(1)$ symmetry.
In fact many works have been carried out in a scheme of THDM with extra $U(1)$ symmetry motivated by several issues such as absence of FCNC~\cite{Ko:2012hd}, neutrino mass~\cite{Cai:2018upp,Bertuzzo:2018ftf,Nomura:2017wxf,Nomura:2017jxb,Camargo:2018uzw,Nomura:2017ohi}, flavor physics~\cite{Ko:2019tts,Ko:2017quv,DelleRose:2017xil,Crivellin:2015lwa,Bian:2017rpg}, dark matter(DM)~\cite{Camargo:2019ukv,Ko:2015fxa,Ko:2014uka,Nomura:2019vqc,Correia:2019pnn,Correia:2019woz} and collider physics~\cite{Camargo:2018klg,Ko:2013zsa,Nomura:2017lsn}.
Also extra $U(1)$ could be originated from string theory~\cite{Olguin-Trejo:2019hxk}.

In this work, we construct a model based on an extra $U(1)_X$ gauge symmetry which can realize lepton specific (type-X) THDM. 
The type-X THDM is one of the interesting scenario in THDM in which one Higgs doublet only couples to quarks while the others only couples to leptons~\cite{Cao:2009as}.
Interestingly one can obtain sizable contribution to muon anomalous magnetic moment (muon $g-2$) from the structure of Yukawa coupling
where the deviation from the SM prediction is~\cite{Davier:2010nc, Hagiwara:2011af,Davier:2017zfy,Davier:2019can};
\begin{align}
\Delta a_\mu=(26.1\pm8)\times10^{-10},\label{eq:damu}.
\end{align}
It is the $3.3\sigma$ deviation with a positive value, and recent theoretical analysis further indicates 3.7$\sigma$ deviation~\cite{Keshavarzi:2018mgv}.
Moreover, several upcoming experiments such as Fermilab E989 \cite{e989} and J-PARC E34 \cite{jpark}
will provide the result with more precision in future.
To explain the discrepancy a lot of studies have been carried out within type-X~\cite{Broggio:2014mna,Wang:2018hnw,Cherchiglia:2017uwv,Abe:2015oca,Wang:2014sda,Chun:2016hzs,Li:2018aov,Chun:2019oix}, muon specific~\cite{Abe:2017jqo} and general (type-III) THDM~\cite{Ilisie:2015tra,Benbrik:2015evd}. 
We then investigate muon $g-2$ in our model taking into account constraints from the SM Higgs measurements.
In addition, we introduce a scalar dark matter (DM) candidate in our model which is stabilized by discrete $Z_2$ symmetry and its interaction with muon can also contribute to muon $g-2$.
The relic density of DM is estimated to search for parameters accommodating with the observed value imposing constraint from direct detection experiments.
We also discuss possibility of indirect detection experiments.

This paper is organized as follows.
In Sec. II, we introduce our model and formulate mass spectrum and interactions.
In Sec. III, we discuss phenomenology of the model such as constraints from scalar potential, muon $g-2$, 
$Z'$ boson production at the LHC, and dark matter physics.
 Finally we give summary and discussion.

\section{Model setup}

\begin{table}[t!]
  \centering
  \begin{tabular}{|c|c|c|c|c|c|c|c|c|c|c|c|c|c|} \hline
   Fields  & ~$Q_L$~ & ~$u_R$~ & ~$d_R$~ & ~$L_L$~ & ~$e_R$~ & $~H_1$~ & ~$H_2$~ & ~$E_L$ ~& ~$E_R$~ & ~$N_R$~ & ~$\phi$~ & ~$\chi$~ & ~$\chi'$~  \\ 
    \hline
    $SU(3)_{C}$ & $\bm{3}$ & $\bm{3}$ & $\bm{3}$ & $\bm{1}$ & $\bm{1}$ & $\bm{1}$ & $\bm{1}$ & $\bm{1}$ & $\bm{1}$ & $\bm{1}$ & $\bm{1}$ & $\bm{1}$ & $\bm{1}$ \\
    \hline
    $SU(2)_L$ & $\bm{2}$ & $\bm{1}$ & $\bm{1}$ & $\bm{2}$ & $\bm{1}$ & $\bm{2}$ & $\bm{2}$ & $\bm{1}$ & $\bm{1}$ & $\bm{1}$ & $\bm{1}$ & $\bm{1}$ & $\bm{1}$ \\
    \hline
    $U(1)_Y$ & $\frac16$ & $\frac23$ & $-\frac13$ & $-\frac12$ & $-1$ & $\frac12$ & $\frac12$ & $-1$ & $-1$ & $0$ & $0$ & $0$ & $0$ \\
    \hline
    $U(1)_X$ & $0$ & $1$ & $-1$ & $0$ & $0$ & $0$ & $1$ & $0$ & $-1$ & $1$ & $-1$ & $1$ & $0$ \\
    \hline
    $Z_2$ & $+$ & $+$ & $+$ & $+$ & $+$ & $+$ & $+$ & $-$ & $-$ & $+$ & $+$ & $-$ & $-$ \\ \hline
  \end{tabular}
  \caption{ 
Charge assignments of field contents under $SU(3)_C\times SU(2)_L\times U(1)_Y\times U(1)_{X} \times Z_2$.}
  \label{tab:table1}
\end{table}

In this section, we introduce our model and formulate mass spectrum and interactions. This model has extra $U(1)_X$ gauge symmetry and exotic charged leptons $E_{R(L)}$ with $U(1)_X$ charge $-1(0)$ are introduced to cancel gauge anomalies.  
In scalar sector, we introduce two Higgs doublets $H_{1}$ and $H_2$ whose $U(1)_X$ charges are $0$ and $1$ respectively,  
and complex SM singlet scalars $\phi$, $\chi$ and $\chi'$ with $U(1)_X$ charge $-1$, $1$ and $0$.
We also impose $Z_2$ parity where $E_{L(R)}$, $\chi$ and $\chi'$ are odd and the other fields are even, and neutral scalar $\chi$ and $\chi'$ can be our DM candidate~\cite{Baek:2018wuo}.
Here we consider two DM candidates $\chi$ and $\chi'$ where the former has gauge interaction associated with $U(1)_X$ and the other is gauge singlet.
The full charge assignment of fields are summarized in Table~\ref{tab:table1}.  
Scalar fields in our model are written as 
\begin{equation}
H_{i} = \begin{pmatrix} \phi_{i}^+ \\ \frac{1}{\sqrt{2}} (v_i + h_i + i a_i) \end{pmatrix}, \quad \phi = \frac{1}{\sqrt{2}} (\eta + \phi_R + i \phi_I), \quad 
 \chi(\chi') = \frac{1}{\sqrt{2}} \left(\chi_R(\chi'_R) + i \chi_I(\chi'_I) \right),
\end{equation}
where $i =1,2$, and $v_i$ and $\eta$ are VEVs of corresponding fields.
We require $\chi(\chi')$ not to develop VEV so that $Z_2$ symmetry is not broken.

Here we show that our fermion contents satisfy the gauge and gravity anomaly free condition as follows
\begin{align}
& U(1)_X \times [SU(3)_c]^2 \,\,\,\,\, :\,\,\,\,\,  Q_X^{d_R} + Q_{X}^{d_R} = 1 - 1 = 0 \nonumber , \\
&   U(1)_X \times [U(1)_Y]^2   \,\,\,\,\, :\,\,\,\,\, 3 \left( \frac{2}{3} \right)^2 Q_{X}^{u_R} + 3 \left( \frac{-1}{3} \right)^2 Q_{X}^{d_R} +  \left( -1 \right)^2 Q_{X}^{E_R} =  \frac{4}{3} - \frac{1}{3} - 1 = 0, \nonumber \\
&  [U(1)_X]^2\times U(1)_Y  \,\,\,\,\, :\,\,\,\,\, 3 \left( \frac{2}{3} \right) (Q_{X}^{u_R})^2 +  3 \left( \frac{-1}{3} \right) (Q_{X}^{d_R})^2 + (-1)  (Q_{X}^{E_R})^2 = 2 -1 -1 =0, \nonumber \\
& \nonumber  [U(1)_X]^3  \,\,\,\,\, :\,\,\,\,\, 3(Q_{X}^{u_R})^3 + 3(Q_{X}^{d_R})^3 + (Q_{X}^{E_R})^3 + (Q_{X}^{N_R})^3 = 3 -3  - 1 +1 = 0 ,\nonumber \\
& U(1)'\times [\textrm{grav}]^2  \,\,\,\,\, :\,\,\,\,\, 3(Q_{X}^{u_R}) + 3(Q_{X}^{d_R}) + (Q_{X}^{E_R}) + (Q_{X}^{N_R})  =  3 -3 -1 +1 =0,
\end{align}
where $Q_X^{f_{SM}}$ is the $U(1)_X$ charge of the SM fermion $f_{SM}$, 
and the condition associated with $SU(2)_L$ is the same as the SM since $SU(2)_L$ doublet fermions do not have $U(1)_X$ charge.
This structure of anomaly cancellation is similar to right-handed fermion specific $U(1)$ case~\cite{Ko:2012hd, Nomura:2017tih,Nomura:2016emz} where the extra charged lepton play a role of right-handed charged lepton in our case.

\subsection{Scalar sector \label{sec:scalar}}

Here we discuss scalar sector in the model formulating mass spectrum and corresponding mass eigenstates.
 The scalar potential is given by
\begin{align}
V = & m_1^2H_1^{\dagger}H_1 + m^2_2H_2^{\dagger}H_2 + m_{\phi}^2\phi^{*}\phi+M_{\chi}^2\chi^{*}\chi +M_{\chi'}^2\chi'^{*}\chi'  - \mu(H_1^{\dagger}H_2\phi+h.c.) + \lambda_1(H_1^{\dagger}H_1)^2 \nonumber\\ 
& + \lambda_2(H_2^{\dagger}H_2)^2   +\lambda_3(H_1^{\dagger}H_1)(H_2^{\dagger}H_2) + \lambda_4(H_1^{\dagger}H_2)(H_2^{\dagger}H_1) + \lambda_{\phi}(\phi^{*}\phi)^2 + \lambda_{\phi H_1}(H_1^{\dagger}H_1)(\phi^{*}\phi) \nonumber \\ 
& + \lambda_{\phi H_2}(H_2^{\dagger}H_2)(\phi^{\dagger}\phi) + \lambda_{\chi_\phi}(\chi^{*}\chi)(\phi^{*}\phi) + \lambda_{\chi H_1}(\chi^{*}\chi)(H_1^{\dagger}H_1) + \lambda_{\chi H_2}(\chi^{*}\chi)(H_2^{\dagger}H_2) \nonumber \\
& + \lambda_{\chi'_\phi}(\chi'^{*}\chi')(\phi^{*}\phi) + \lambda_{\chi' H_1}(\chi'^{*}\chi')(H_1^{\dagger}H_1) + \lambda_{\chi' H_2}(\chi'^{*}\chi')(H_2^{\dagger}H_2) \nonumber \\ 
&  + \lambda_\chi (\chi^* \chi)^2 + \lambda_{\chi'} (\chi'^* \chi')^2 + \mu_\chi (\chi \chi' \phi + h.c. ),
\end{align} 
where we take the couplings to be real for simplicity.
In addition we require invariance under phase transformation $\chi \to e^{i \theta_\chi} \chi$ and $\chi' \to e^{i \theta_{\chi'}} \chi'$ to simplify the scalar potential,  which is softly broken by the last term of the potential.
The VEVs can be obtained by solving the condition $\partial V/\partial v_1 = \partial V/\partial v_2 = \partial V/\partial \eta =0$.
From the condition, we require the VEVs and parameters to satisfy
\begin{eqnarray}
m_1^2v_1 +v_1^3\lambda_1 + \frac{1}{2}v_1v_2^2\lambda_3 + \frac{1}{2}v_1v_2^2\lambda_4 +\frac{1}{2} v_1\eta^2\lambda_{\phi H_1} +\frac{1}{\sqrt{2}} v_2\eta\mu=0 \nonumber \\
m_2^2v_2 + v_2^3\lambda_2 + \frac{1}{2}v_1^2v_2\lambda_3 +\frac{1}{2}v_1^2v_2\lambda_4 + \frac{1}{2}{}v_2\eta^2\lambda_{\phi H_2} + \frac{1}{\sqrt{2}} v_1\eta\mu=0 \nonumber \\
m_{\phi}^2\eta + \eta^3\lambda_{\phi} +\frac{1}{2} v_1^2\eta\lambda_{\phi H_1} +\frac{1}{2} v_2^2\eta\lambda_{\phi H_2} + \frac{1}{\sqrt{2}} v_1v_2\mu=0.
\end{eqnarray}
Also to obtain vanishing VEV of $\chi (\chi')$, we require $M_{\chi(\chi')}^2$ and couplings associated with $\chi(\chi')$ to be positive.

After spontaneous symmetry breaking, we obtain mass matrix for charged scalar such that
\begin{equation}
\mathcal{L} \supset \begin{pmatrix} \phi_1^- \\ \phi_2^- \end{pmatrix}^T \left( \frac{\eta \mu}{\sqrt2 v_1 v_2} - \frac{\lambda_4}{2} \right) 
\begin{pmatrix} v_2^2 & - v_1 v_2 \\ - v_1 v_2 & v_2^2 \end{pmatrix}
\begin{pmatrix} \phi_1^+ \\ \phi_2^+ \end{pmatrix}.
\end{equation}
The mass matrix can be diagonalized as in the THDM and mass eigenstates are 
\begin{equation}
\left( \begin{array}{c} G^\pm \\ H^\pm \end{array} \right)
= \left( \begin{array}{cc} \cos \beta & - \sin \beta \\ \sin \beta & \cos \beta \end{array} \right) 
\left( \begin{array}{c} \phi_1^\pm \\ \phi_2^\pm \end{array} \right),
\label{Eq:chargedMES}
\end{equation}
where $\tan \beta = v_2/v_1$, $G^\pm$ is Nambu-Goldstone(NG) boson absorbed by $W^\pm$ and $H^\pm$ is physical charged Higgs boson. 
The mass of charged Higgs boson is given by
\begin{equation}
m_{H^\pm}^2 =  \frac{\mu \eta}{\sqrt{2} \sin \beta \cos \beta} - \frac{\lambda_4}{2} v^2, 
\end{equation}
where $v = \sqrt{v_1^2 + v_2^2}$.

The mass matrix for $Z_2$ even and CP odd scalar bosons is obtained as 
\begin{align}
\mathcal{L} \supset
\frac12 \left(\begin{array}{ccc}
a_1 \\
a_2 \\
\phi_I\\
\end{array}
\right)^T
\left(
\begin{array}{ccc}
 \frac{\eta  \mu  \text{v2}}{\sqrt{2} v_1} & -\frac{\eta  \mu }{\sqrt{2}} & -\frac{\mu v_2}{\sqrt{2}} \\
 - \frac{\eta  \mu }{\sqrt{2}} &  \frac{\eta  \mu  v_1}{\sqrt{2} v_2} &  \frac{\mu  v_1}{\sqrt{2}} \\
 - \frac{\mu  v_2}{\sqrt{2}} &  \frac{\mu  v_1}{\sqrt{2}} &  \frac{\mu  v_1 v_2}{\sqrt{2} \eta } \\
\end{array}
\right)
\left(\begin{array}{ccc}
a_1 \\
a_2 \\
\phi_I\\
\end{array}
\right).
\end{align}
We can diagonalize the mass matrix by rotating the basis as follows:
\begin{equation}
\left(\begin{array}{ccc}
a_1 \\
a_2 \\
\phi_I\\
\end{array}
\right)=
\left(
\begin{array}{ccc}
 \frac{\text{v1}}{\sqrt{v_1^2+\text{v2}^2}} & -\frac{\eta  v_2}{\sqrt{\eta^2 v_1^2+\text{v1}^2 v_2^2+\eta^2 v_2^2}} & \frac{v_1}{\sqrt{\eta^2 + v_1^2}} \\
 \frac{v_2}{\sqrt{v_1^2 + v_2^2}} & \frac{\eta  v_1}{\sqrt{\eta^2 v_1^2 + v_1^2 v_2^2+\eta^2 v_2^2}} & 0 \\
 0 & \frac{v_1 v_2}{\sqrt{\eta ^2 v_1^2 + v_1^2 v_2^2+\eta ^2 v_2^2}} & \frac{\eta }{\sqrt{\eta^2 + v_1^2}} \\
\end{array}
\right)
\left(\begin{array}{ccc}
G^0_{1} \\
A^0 \\
G^0_{2} \\
\end{array}
\right),
\label{eq:CP-Odd}
\end{equation} 
where $G_1^0$ and $G_{2}^0$ are massless NG bosons and these degrees of freedom are absorbed by $Z$ and $Z'$ bosons.
The physical CP-odd scalar boson $A^0$ has non-zero mass of 
\begin{equation}
m^2_{A^0} = \frac{\mu ( \eta^2 + c_\beta s_\beta v^2)}{\sqrt{2} c_\beta s_\beta \eta }.
\end{equation}
We thus find that $A^0$ becomes massless in the limit of $\mu \to 0$.

The $Z_2$ even and CP-even scalar sector has three physical degrees of freedom $\{h_1, h_2, \phi_R \}$ and the mass matrix is given by
\begin{equation}
\mathcal{L} \supset \frac{1}{2}
\left( \begin{array}{c} h_1 \\ h_2 \\ \phi_R \end{array} \right)^T 
\left(
\begin{array}{ccc}
 2 \lambda_1 v_1^2 + \frac{\eta  \mu  v_2}{\sqrt{2} v_1} & \lambda _3 v_1 v_2+\lambda_4 v_1 v_2 - \frac{\eta  \mu }{\sqrt{2}} & \eta \lambda_{\phi H1} v_1 - \frac{\mu  v_2}{\sqrt{2}} \\
 \lambda_3 v_1 v_2+\lambda_4 v_1 v_2 - \frac{\eta \mu }{\sqrt{2}} & 2 \lambda_2v_2^2 + \frac{\eta \mu  v_1}{\sqrt{2} v_2} & \eta\lambda_{\phi H2}v_2 -\frac{\mu  v_1}{\sqrt{2}}  \\
 \eta  \lambda_{\phi H1} v_1 - \frac{\mu v_2}{\sqrt{2}} & \eta \lambda_{\phi H2} v_2 - \frac{\mu v_1}{\sqrt{2}} & 2 \eta^2 \lambda_{\phi} + \frac{\mu  v_1 v_2}{\sqrt{2} \eta } \\
\end{array}
\right) 
\left( \begin{array}{c} h_1 \\ h_2 \\ \phi_R \end{array} \right).
\end{equation}
This mass matrix can be diagonalized by an orthogonal matrix $R$ with three Euler parameters $\{ \alpha_1, \alpha_2, \alpha_3 \}$ which is written as
\begin{equation}
\\ R(\alpha_1,\alpha_2,\alpha_3)=\left(\begin{array}{ccc}
c_{\alpha_1}c_{\alpha_2} & - s_{\alpha_1}c_{\alpha_2} & s_{\alpha_2} \\
- c_{\alpha_1}s_{\alpha_2}s_{\alpha_3} + s_{\alpha_1}c_{\alpha_3} & c_{\alpha_1}c_{\alpha_3}+s_{\alpha_1}s_{\alpha_2}s_{\alpha_3} & c_{\alpha_2}s_{\alpha_3}\\
-c_{\alpha_1}s_{\alpha_2}c_{\alpha_3}-s_{\alpha_1}s_{\alpha_3} & - c_{\alpha_1}s_{\alpha_3} + s_{\alpha_1}s_{\alpha_2}c_{\alpha_3}& c_{\alpha_2}c_{\alpha_3}\\
\end{array}\right)
\end{equation} 
and mass eigenstates are obtained such that
\begin{equation}
\left( \begin{array}{c} h_1 \\ h_2 \\ \phi_R \end{array} \right) = R_{ij} \left( \begin{array}{c} H^0 \\ h^0 \\ \xi^0 \end{array} \right)_j.
\label{Eq:CP-even}
\end{equation}

We write parameters in scalar potential 
$\{ m_1,m_2,\mu,\lambda_1,\lambda_2,\lambda_3, \lambda_4, \lambda_{\phi},\lambda_{\phi H_1},\lambda_{\phi H2} \}$
by physical masses and VEVs such that 
\begin{align}
\mu &= \frac{\sqrt{2} m_{A^0}^2\eta}{(v^2\sin{\beta}\cos{\beta}+\eta^2\cot{\beta}+\eta^2\tan{\beta})}\\
\lambda_4 & =  \frac{2}{v^2} \left(\frac{\eta\mu}{\sqrt{2}\sin{\beta}\cos{\beta}} - m_{H^2_{\pm}} \right)\\
\lambda_1 &= \frac{2 m_{H^0}^2 R_{11}^2 v_1+2 m_{h^0}^2 R_{12}^2 v_1+2 m_{\xi^0}^2 R_{13}^2 v_1 - \sqrt{2}\eta \mu v_2}{4v_1^3} \\
\lambda_2 &= \frac{2 m_{H^0}^2 R_{21}^2 v_2+2 m_{h^0}^2 R_{22}^2 v_{2}+2m_{\xi^0}^2 R_{23}^2 v_{2} - \sqrt{2}\eta\mu v_1}{4 v_2^3} \\
\lambda_3 &= \frac{2 m_{H^0}^2R_{11}R_{21} + 2m_{h_0}^2R_{12}R_{22} + 2m_{\xi^0}^2R_{13}R_{23}-2v_1v_2\lambda_4 + \sqrt{2}\eta\mu}{2v_1v_2}\\
\lambda_{\phi} &=\frac{2m_{H^0}^2R_{31}^2\eta + 2m_{h^0}^2R_{32}^2\eta + 2m_{\xi^0}^2R_{33}^2\eta - \sqrt{2}v_1v_2\mu}{4\eta^3}\\
\lambda_{\phi H_1} &=\frac{2m_{H^0}^2R_{11}R_{31}+2m_{h^0}^2R_{12}R_{32} + 2m_{\xi^0}^2R_{13}R_{33}+\sqrt{2}v_2\mu}{2v_1\eta}\\
\lambda_{\phi H_2} &=\frac{2m_{H^0}^2R_{21}R_{31} + 2m_{h^0}^2R_{22}R_{32} + 2m_{\xi^0}^2R_{23}R_{33}+\sqrt{2}v_1\mu}{2v_2\eta}.
\end{align}

Here we formulate masses of $Z_2$ odd scalar fields $\chi_{R(I)}[\chi'_{R(I)}]$. 
For simplicity we assume $\mu_\chi \eta \ll M_\chi^2$ and ignore $\chi$-$\chi'$ mixing.
Then masse eigenvalues of them are given by
\begin{align}
m_{\chi_R}^2 \simeq m_{\chi_I}^2  & \simeq  M^2_{\chi}  + \frac{v^2}{2} ( \lambda_{\chi H_1}  \cos^2 \beta + \lambda_{\chi H_2 } \sin^2 \beta)  + \frac{1}{2}  \lambda_{\chi \phi } \eta^2 , \\
m_{\chi'_R}^2 \simeq m_{\chi'_I}^2  & \simeq  M^2_{\chi'}  + \frac{v^2}{2} ( \lambda_{\chi' H_1}  \cos^2 \beta + \lambda_{\chi' H_2 } \sin^2 \beta)  + \frac{1}{2}  \lambda_{\chi' \phi } \eta^2 , 
\end{align}
where the real and imaginary part of $\chi (\chi')$ have the same mass, and we write them as $m_\chi$ and $m_{\chi'}$.
Here the mass degeneracy of real and imaginary part is due to the requirement of invariance under phase transformation and smallness of $\mu_\chi$ parameter in the scalar potential as we assumed above.
Thus our DM is identified as complex scalar bosons.

\subsection{Yukawa interactions}

%%%%%%%%%%%%%%%%%%%%%%%%%%%%%%%%%%%%%%%%%%%%%%%%%%%%%%%%%%%%%%%%%%%%%%
\begin{table}[t]
  \centering
  \begin{tabular}{|c|c|c|c|c|} \hline
~~$\Phi$~~ & ~~$H^0$~~ & ~~$h^0$~~ & ~~$\xi^0$~~ & ~~$A^0$~~ \\ \hline
 $y_\Phi^u$ & $\frac{R_{21}}{\sin \beta}$ & $\frac{R_{22}}{\sin \beta}$ &  $\frac{R_{23}}{\sin \beta}$ & $- \frac{O_{22}}{\sin \beta}$ \\ \hline
  $y_\Phi^d$ & $\frac{R_{21}}{\sin \beta}$ & $\frac{R_{22}}{\sin \beta}$ &  $\frac{R_{23}}{\sin \beta}$ & $ \frac{O_{22}}{\sin \beta}$ \\ \hline
  $y_\Phi^e$ & $\frac{R_{11}}{\cos \beta}$ & $\frac{R_{12}}{\cos \beta}$ &  $\frac{R_{13}}{\cos \beta}$ & $\frac{O_{12}}{\cos \beta}$ \\ \hline
  $y_\Phi^\nu$ & $\frac{R_{11}}{\cos \beta}$ & $\frac{R_{12}}{\cos \beta}$ &  $\frac{R_{13}}{\cos \beta}$ & $-\frac{O_{12}}{\cos \beta}$ \\ \hline
   $y_\Phi^E$ & $\frac{R_{31} v}{\eta}$ & $\frac{R_{32} v}{\eta}$ &  $\frac{R_{33} v}{\eta}$ & $\frac{O_{32} v}{\eta}$ \\ \hline
    \end{tabular} 
    \caption{The mixing factors associated with Yukawa interactions in Eq.~(\ref{Eq:Yukawa}).}
      \label{tab:table2}
\end{table}
%%%%%%%%%%%%%%%%%%%%%%%%%%%%%%%%%%%%%%%%%%%%%%%%%%%%%%%%%%%%%%%%%%%%%%

The Yukawa interactions in our model are controlled by $U(1)_X$ gauge symmetry, 
and one obtains lepton specific (type-X) structure for two Higgs doublet scalars and terms associated with exotic charged leptons:
\begin{align}
-\mathcal{L}_Y = & y^{u} \overline{Q}_{L}\widetilde{H}_2 u_{R} + y^{d} \overline{Q}_{L}H_2 d_{R} + y^{e} \overline{L}_{L}H_1 e_{R} + y^{\nu} \overline{L}_{L}\widetilde{H}_2{N}_{R}  \nonumber \\
& + y^E \phi^* \overline{E}_LE_R + Y^{\chi'}\chi' \overline{E}_Le_R + h.c. \, ,
\label{Eq:Yukawa0}
\end{align}
where we omit flavor indices.
We can derive the SM fermion masses the same as the THDM.
In addition the masses of exotic leptons $E$ is given by
\begin{equation}
m_{E_a} = \frac{y^E_a \eta}{\sqrt{2}}.
\end{equation}
Then rewriting scalar fields by mass eigenstates the Yukawa interactions become  
\begin{align}
\mathcal{L}_Y = & - \sum_{f=u,d,e,E} \left( \frac{m_f}{v} y^f_{h^0} \bar f f h^0 + \frac{m_f}{v} y^f_{H^0} \bar f f H^0 + i \frac{m_f}{v} y^f_{A^0} \bar f \gamma_5 f A^0 \right) \nonumber \\
& + \left[ \frac{\sqrt{2} V_{ud}}{v} \bar u ( m_u \cot \beta P_L - m_d \cot \beta P_R ) d H^+ - \frac{\sqrt{2} m_\ell }{v} \tan \beta \bar \nu_L e_R H^+ + h.c. \right],
\label{Eq:Yukawa}
\end{align}
where $V_{ud}$ indicates an element of CKM matrix.
The coefficients associated with neutral scalar bosons, $y_\Phi^f$, are summarized in Table.~\ref{tab:table2} while interactions associated with charged Higgs are the same as the type-X THDM.
In our model neutrino mass is generated as Dirac type and mass matrix is simply given by $m_\nu = y^\nu v_2/\sqrt{2}$ from Yukawa interaction Eq.~(\ref{Eq:Yukawa0}).
Note that neutrino $\nu_L$ in Eq.~(\ref{Eq:Yukawa}) corresponds to flavor eigenstate.

\subsection{Gauge sector \label{sec:Zp}}

Here we formulate mass eigenvalues and corresponding eigenstates in our gauge sector~\footnote{In our analysis we ignore kinetic mixing between $U(1)_Y$ and $U(1)_X$ gauge fields assuming its effect is negligibly small.}. 
After symmetry breaking gauge bosons obtain masses from kinetic term of scalar fields
\begin{align}
& \mathcal{L}_K = (D_\mu H_1)^\dagger (D^\mu H_1) + (D_\mu H_2)^\dagger (D^\mu H_2) + (D_\mu \phi)^\dagger (D^\mu \phi), \\
& D_\mu H_1 = (\partial_{\mu} + ig\frac{\tau^a}{2}W^a_{\mu} + \frac{1}{2} ig^{\prime}B_{\mu}) H_1, \\
& D_\mu H_2 = (\partial_{\mu} + ig\frac{\tau^a}{2}W^a_{\mu} + \frac{1}{2} ig^{\prime}B_{\mu} + ig^{\prime\prime}B^{\prime}_{\mu}) H_2, \\
& D_\mu \phi = (\partial_{\mu}  - ig^{\prime\prime}B^{\prime}_{\mu}) \phi,
\end{align}
where $g$, $g'$ and $g''$ are gauge couplings associated with $SU(2)_L$, $U(1)_Y$ and $U(1)_X$.
The mass of $W$ boson is given by $m_W = g v/2$ with mass eigenstate $W^\pm_\mu = (W^1_\mu \mp i W^2_\mu)$ as in the SM.
On the other hand mass matrix for neutral gauge bosons becomes $3 \times 3$ such that
\begin{equation}
\mathcal{L^{\text{mass}}_{\text{gauge}}}=\frac{1}{8}\left(\begin{array}{ccc}
W^3_{\mu} \\
B_{\mu} \\
B^{\prime}_{\mu}
\end{array}
\right)^T\left(\begin{array}{ccc}
g^2(v_1^2+v_2^2) & -gg^{\prime}(v_1^2+v_2^2) & -2gg^{\prime\prime}v_2^2 \\
-gg^{\prime}(v_1^2+v_2^2) & g^{\prime 2}(v_1^2+v_2^2) & +2g^{\prime}g^{\prime \prime}v_2^2\\
 -2gg^{\prime\prime}v_2^2 & +2g^{\prime}g^{\prime \prime}v_2^2 & 4g^{\prime\prime 2}v_2^2 +4g^{\prime\prime 2}\eta^2
\end{array}\right)
\left(\begin{array}{ccc}
W^{3\mu} \\
B^{\mu} \\
B^{\prime \mu}
\end{array}
\right).
\end{equation}
Rotating $(W^3_\mu, B_\mu)^T$ by Weinberg angle $\theta_W$, we identify massless photon field $A_\mu$ as
\begin{eqnarray}
\left(\begin{array}{ccc}
W^{3\mu}\\
B^{\mu}\\
\end{array}\right)=
\left(\begin{array}{ccc}
c_W & s_W \\
-s_W & c_W\\
\end{array}\right)
\left(\begin{array}{ccc}
\widetilde{Z}^{\mu}\\
A^{\mu}\\
\end{array}\right),
\end{eqnarray}
where $c_W(s_W) = \cos \theta_W (\sin \theta_W)$ whose definition is the same as in the SM.
Then we obtain $2 \times 2$ mass matrix in the basis of $(\tilde Z_\mu, B'_\mu)$ such that
\begin{align}
\mathcal{L}_{\text{gauge}}^{mass}=\frac{1}{2}
\left(\begin{array}{cc}
\widetilde{Z}\\
B^{\prime}
\end{array}\right)^T
\left(\begin{array}{cc}
M_{Z,SM}^2 & -\Delta^2\\
-\Delta^2  & M_{Z^\prime}\\
\end{array}\right)
\left(\begin{array}{cc}
\widetilde{Z}\\
B^{\prime}
\end{array}\right)
\end{align}
where the elements are given by
\begin{align}
M_{Z, SM}^2 = \frac{1}{4} (g^2 + g'^2) v^2, \quad M^2_{Z'} = g''^2 (v_1^2 + \eta^2), \quad \Delta^2 = \frac{2 g''}{\sqrt{g^2 + g'^2}} M^2_{Z, SM} \sin^2 \beta.
\end{align}
The mass eigenvalues are 
\begin{align}
m_Z^2 & =\frac{1}{2}\Big(M_{Z,SM}^2 + M_{Z^{\prime}}^2 -\sqrt{(M_{Z,SM}^2-M_{Z^{\prime}}^2)^2 + 4\Delta^4}\Big), \\
m_{Z^{\prime}}^2 &=\frac{1}{2}\Big(M_{Z,SM}^2 + M_{Z^{\prime}}^2 +\sqrt{(M_{Z,SM}^2-M_{Z^{\prime}}^2)^2 + 4\Delta^4}\Big),
\end{align}
and the mass eigenstates are obtained such that
\begin{align}
& \left(\begin{array}{cc}
Z \\
Z^{\prime}
\end{array}\right)=
\left(\begin{array}{cc}
\cos{\theta_{ZZ^{\prime}}} & \sin{\theta_{ZZ^{\prime}}}\\
-\sin{\theta_{ZZ^{\prime}}} & \cos{\theta_{ZZ^{\prime}}}\
\end{array}\right)\left(\begin{array}{cc}
\widetilde{Z} \\
B^{\prime}
\end{array}\right)\\
& \tan{2\theta_{ZZ^{\prime}}}=\frac{2\Delta^2}{M_{Z^\prime}^2 -M_{Z,SM}^2}.
\end{align}
The mixing between $Z$ and $Z'$ is sufficiently small in our parameter region of interest and 
we ignore the effect of the mixing in the following analysis.

The gauge interactions among $Z'$ and fermions are given by
\begin{align}
\mathcal{L} \supset g'' Z'_\mu (\bar u_R \gamma^\mu u_R  - \bar d_R \gamma^\mu d_R - \bar E_R \gamma^\mu E_R + \bar N_R \gamma^\mu N_R).
\end{align}
We also obtain $Z'$-scalar-scalar gauge interactions such that
\begin{align}
\mathcal{L} \supset & \ i g'' c^2_\beta Z'_\mu (\phi_2^- \partial^\mu \phi_2^+ - \phi_2^+ \partial^\mu \phi_2^-) +
 g'' \frac{\eta v c_\beta R_{21} - v^2 c_\beta s_\beta R_{31}}{\sqrt{\eta^2 v^2 + v^4 s^2_\beta c^2_\beta}} Z'_\mu (A^0 \partial^\mu H^0 - H^0 \partial^\mu A^0) \nonumber \\
& +  g'' \frac{\eta v c_\beta R_{22} - v^2 c_\beta s_\beta R_{32}}{\sqrt{\eta^2 v^2 + v^4 s^2_\beta c^2_\beta}} Z'_\mu (A^0 \partial^\mu h^0 - h^0 \partial^\mu A^0) \nonumber \\
 & +  g'' \frac{\eta v c_\beta R_{23} - v^2 c_\beta s_\beta R_{33}}{\sqrt{\eta^2 v^2 + v^4 s^2_\beta c^2_\beta}} Z'_\mu (A^0 \partial^\mu \xi^0 - \xi^0 \partial^\mu A^0),
\end{align}
where $c_\beta (s_\beta) = \cos \beta (\sin \beta)$.
In addition the $h^0 VV$ and $H^0 VV$ interactions are given by
\begin{align}
\mathcal{L} \supset & \frac{1}{2} \frac{m_Z^2}{v} (c_\beta R_{12} + s_\beta R_{22}) h^0 Z_\mu Z^\mu + \frac{m_W^2}{v} (c_\beta R_{12} + s_\beta R_{22}) h^0 W^+_\mu W^{-\mu} \nonumber \\
&+ \frac{1}{2} \frac{m_Z^2}{v} (c_\beta R_{11} + s_\beta R_{21}) H^0 Z_\mu Z^\mu + \frac{m_W^2}{v} (c_\beta R_{11} + s_\beta R_{21}) H^0 W^+_\mu W^{-\mu}.
\end{align}
Note that we reproduce THDM interaction in the limit of $\alpha_1 \to \alpha$, $\alpha_2 \to 0$ and $\alpha_3 \to 0$ as 
$c_\beta R_{12} + s_\beta R_{22} \to \sin (\beta - \alpha)$ and $c_\beta R_{11} + s_\beta R_{21} \to \cos (\beta - \alpha)$.

\section{Constraints and phenomenology}

In this section, we discuss experimental constraints and phenomenologies in the model.
We first investigate constraints from Higgs sector such as stability and perturbativity bound in the potential, in order to search for allowed parameter region.
Then muon anomalous magnetic moment is estimated applying the allowed parameter sets.
We also explore collider phenomenology and dark matter physics.

\subsection{Constraints from Higgs sector \label{sec:const}}

Here we discuss constraints on our parameters such as neutral scalar mixing $\{ \alpha_1, \alpha_2, \alpha_3 \}$, scalar boson masses and $\tan \beta$ taking into account 
unitarity, stability and perturbativity bounds for the Higgs sector as well as the experimental measurements of SM Higgs coupling strength.
The constraints from unitary and perturbativity are given by~\cite{Bian:2017xzg} 
\begin{align}
& |\lambda_{1,2,3,\phi }| \leq 4 \pi, \quad |\lambda_{\phi H_{1,2}}| \leq 8 \pi, \quad |\lambda_{3} \pm \lambda_4| \leq 8 \pi, \quad  |\lambda_3 + 2 \lambda_4| \leq 8 \pi, \nonumber \\
& \sqrt{|\lambda_3 (\lambda_3 + 2 \lambda_4)|} \leq 8 \pi,  \quad
\left| \lambda_1 + \lambda_2 \pm \sqrt{(\lambda_1 - \lambda_2)^2 + \lambda^2_4} \right| \leq 8 \pi, \quad a_{1,2,3} \leq 8 \pi,
\end{align}
where $a_{1,2,3}$ are the solution of the following equation
\begin{align}
& x^3 - 2x^2 (3 \lambda_1 + 3 \lambda_2 + 2 \lambda_{\phi})  \nonumber \\
& -  x (2 \lambda_{\phi H_1}^2 + 2 \lambda_{\phi H_2}^2 - 36 \lambda_1 \lambda_2 - 24 \lambda_1 \lambda_\phi - 24 \lambda_1 \lambda_\phi + 4 \lambda_3^2 + 4\lambda_3 \lambda_4 + \lambda_4^2 ) \nonumber \\
& + 4 (3 \lambda_{\phi H_1}^2 \lambda_2 - \lambda_{\phi H_1} \lambda_{\phi H_2} (2 \lambda_3 + \lambda_4) + 3 \lambda_{\phi H_2}^2 \lambda_1 + \lambda_{\phi} ((2 \lambda_3 + \lambda_4)^2 - 36 \lambda_1 \lambda_2)) =0.
\end{align}
We also obtain constraints from stability condition for scalar potential such that~\cite{ElKaffas:2006gdt,Grzadkowski:2009bt,Drozd:2014yla}
\begin{align}
& \lambda_{1,2,\phi} > 0, \quad 2 \sqrt{\lambda_1 \lambda_2} + \lambda_3 + \lambda_4 > 0, \nonumber \\
& 2 \sqrt{\lambda_1 \lambda_2} + \lambda_3 > 0, \quad 2 \sqrt{\lambda_1 \lambda_\phi} + \lambda_{\phi H_1} > 0, \quad 2 \sqrt{\lambda_2 \lambda_\phi} +  \lambda_{\phi H_2} > 0, \nonumber \\
& \sqrt{(\lambda_{\phi H_1}^2 - 4 \lambda_1 \lambda_\phi )(\lambda_{\phi H_2}^2 - 4 \lambda_2 \lambda_\phi )} + 2 \lambda_3 \lambda_\phi > \lambda_{\phi H_1 } \lambda_{\phi H_2}, \nonumber \\
& \sqrt{(\lambda_{\phi H_1}^2 - 4 \lambda_1 \lambda_\phi )(\lambda_{\phi H_2}^2 - 4 \lambda_2 \lambda_\phi )} + 2 (\lambda_3 + \lambda_4) \lambda_\phi > \lambda_{\phi H_1 } \lambda_{\phi H_2}.
\end{align}
Note that we do not consider couplings associated with $\chi$ since it does not develop VEV and we just assume that these couplings are positive values and not too large satisfying perturbativity and unitarity condition. 
Furthermore we impose constraint from the SM Higgs coupling measurements as follows
\begin{align}
& 1.22 > \kappa_V > 0.87, \quad 1.26 > \kappa_t > 0.81, \quad 1.45 > \kappa_b > 0.55, \quad 1.36 > \kappa_\tau > 0.70,  \\
& \kappa_V = c_\beta R_{12} + s_\beta R_{22}, \quad \kappa_t = \kappa_b = \frac{R_{22}}{s_\beta}, \quad \kappa_\tau = \frac{R_{12}}{c_\beta},
\end{align}
where we have applied $2 \sigma$ region of observed values in refs.~\cite{ATLAS:2018doi, Sirunyan:2018koj}.

 \begin{figure}[t!]
\includegraphics[width=80mm]{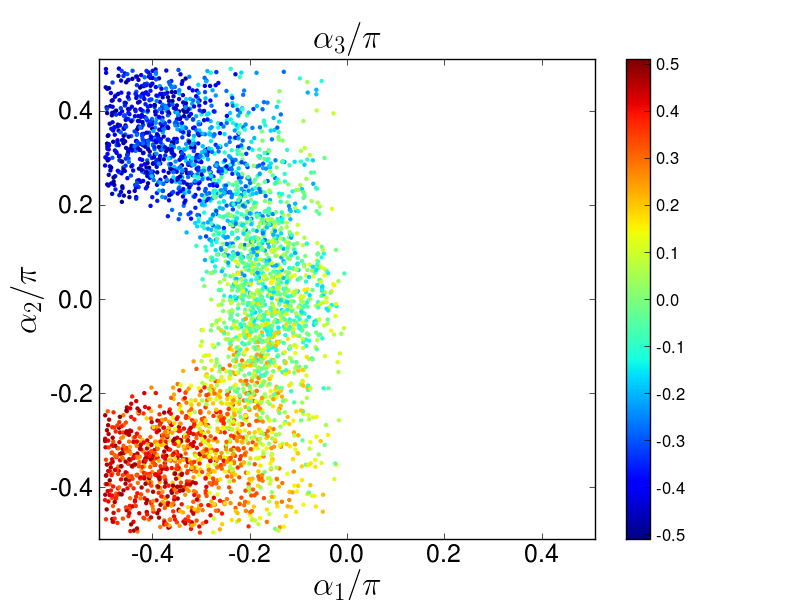}
\includegraphics[width=80mm]{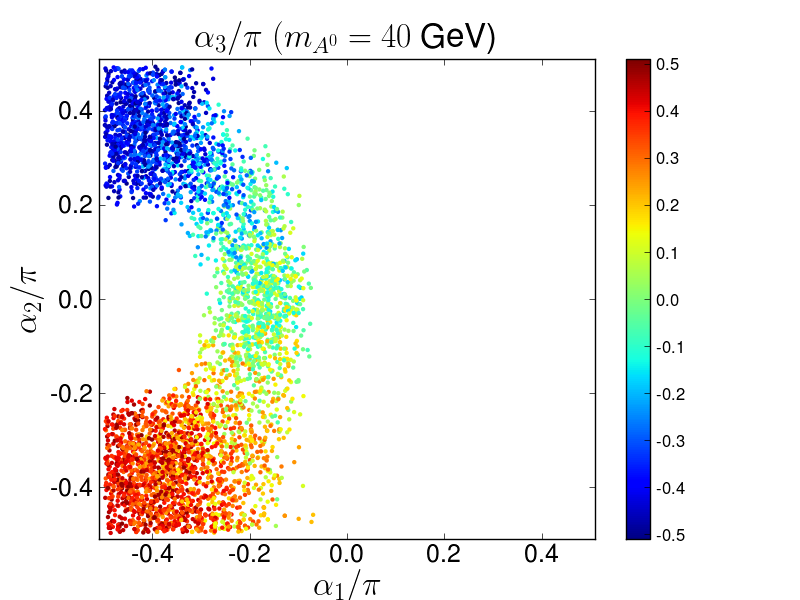} 
\includegraphics[width=80mm]{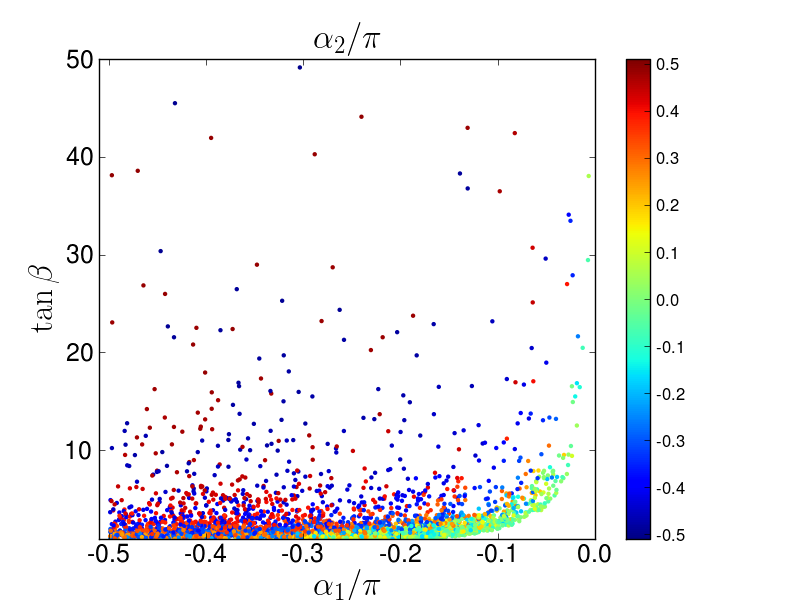}
\includegraphics[width=80mm]{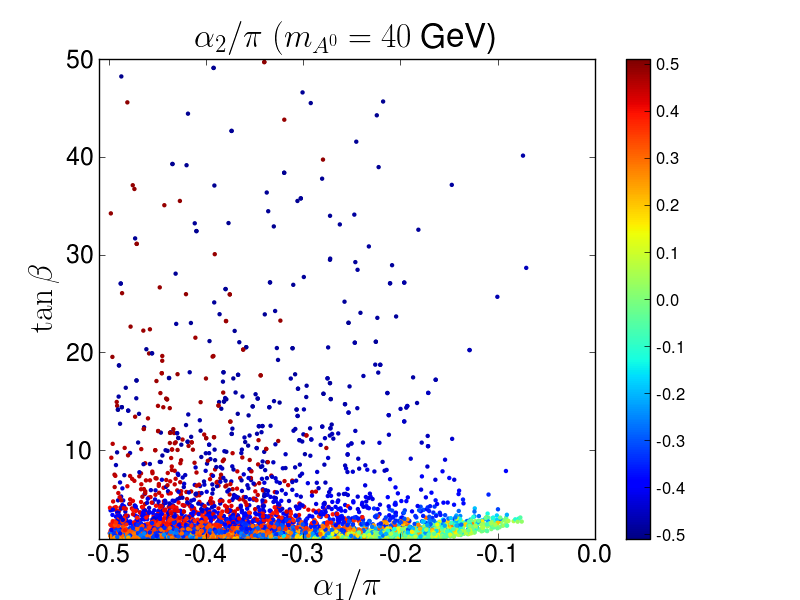}
\includegraphics[width=80mm]{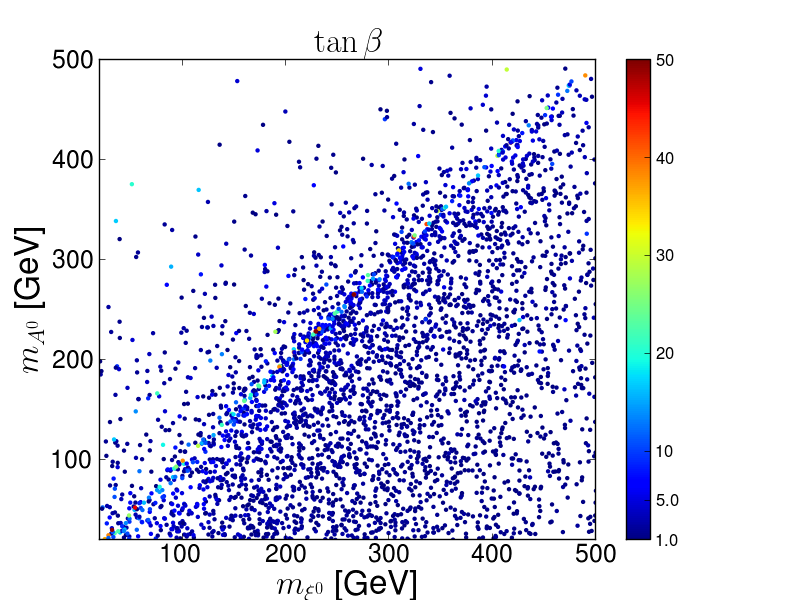}
\includegraphics[width=80mm]{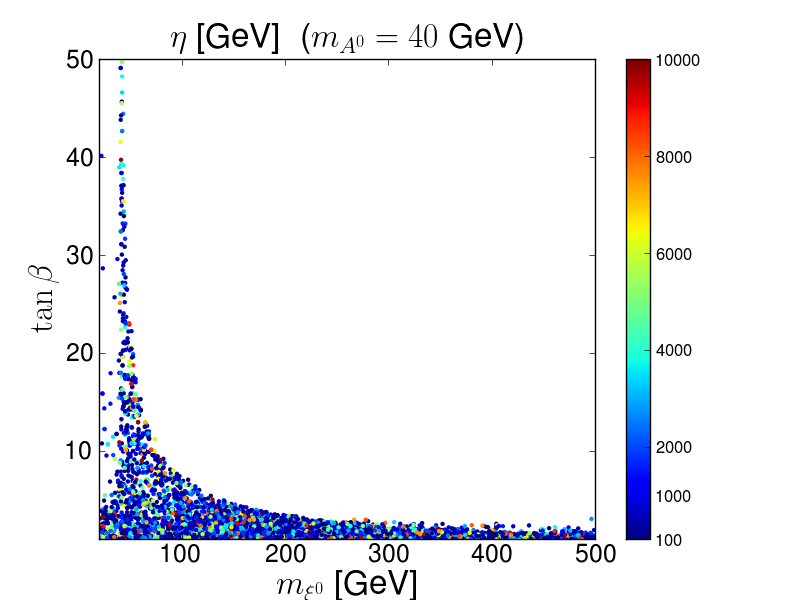}
\caption{The allowed parameter regions where we take $m_{A^0}$ as a scanning parameter in left side plots 
and $m_{A^0} = 40$ GeV is chosen in right side plots. The color gradient corresponds to the values of the parameter indicated by a top label. }
\label{fig:allowed}
\end{figure}

Here we scan out parameters to search for allowed parameter region, such that
\begin{align}
& \alpha_{1,2,3} \in \left[-\frac{\pi}{2}, \frac{\pi}{2} \right], \quad m_{H^0} = m_{H^\pm} \in [200, 500] \ {\rm [GeV]}, \quad m_{A^0, \xi^0} \in [40, 500] \ {\rm [GeV]}, \nonumber \\
& \tan \beta \in [1, 50], \quad \eta \in [100, 10000] \ {\rm [GeV]},
\end{align} 
where we can take range of mixing angle in $[\pi/2, \pi/2]$ without loss of generality.
The allowed parameter regions are shown in Fig.~\ref{fig:allowed} where we take $m_{A^0}$ as a scanning parameter in left side plots 
and $m_{A^0} = 40$ GeV is chosen in right side plots.
We find that relations among mixing angle $\alpha_1 < 0$ and $\alpha_2 = - \alpha_3$ are required to satisfy the constraints.
Furthermore correlations of parameters $|\alpha_2| \sim \pi/2$ and $m_{A^0} \sim m_{\xi^0}$ are preferred to obtain large $\tan \beta$.
On the other hand value of $\eta$ is not strongly constrained and does not correlate with the other parameters.
We can thus take $\eta$ as almost free parameter.

\subsection{Muon $g-2$}

Here we estimate muon $g-2$ in our model. 
Firstly we have contributions from loop diagrams with $Z_2$ even scalar bosons at one- and two-loop level.
The two-loop Barr-Zee type diagrams can provide sizable contributions to muon $g-2$ and the formula is given in refs.~\cite{Ilisie:2015tra,Barr:1990vd}.
We find that sum of contributions to muon $g-2$ from loop diagrams associated with $\phi = \{h^0, H^0, \xi^0, A^0\}$ is at most $\mathcal{O}(10^{-10})$
when we apply allowed parameter region satisfying constraints discussed in previous subsection.
This behavior is due to the negative contribution from two loop diagram associated with $\xi^0$. 
We thus need the other contribution to explain muon $g-2$ in the model.

\begin{figure}[tb]
\includegraphics[width=80mm]{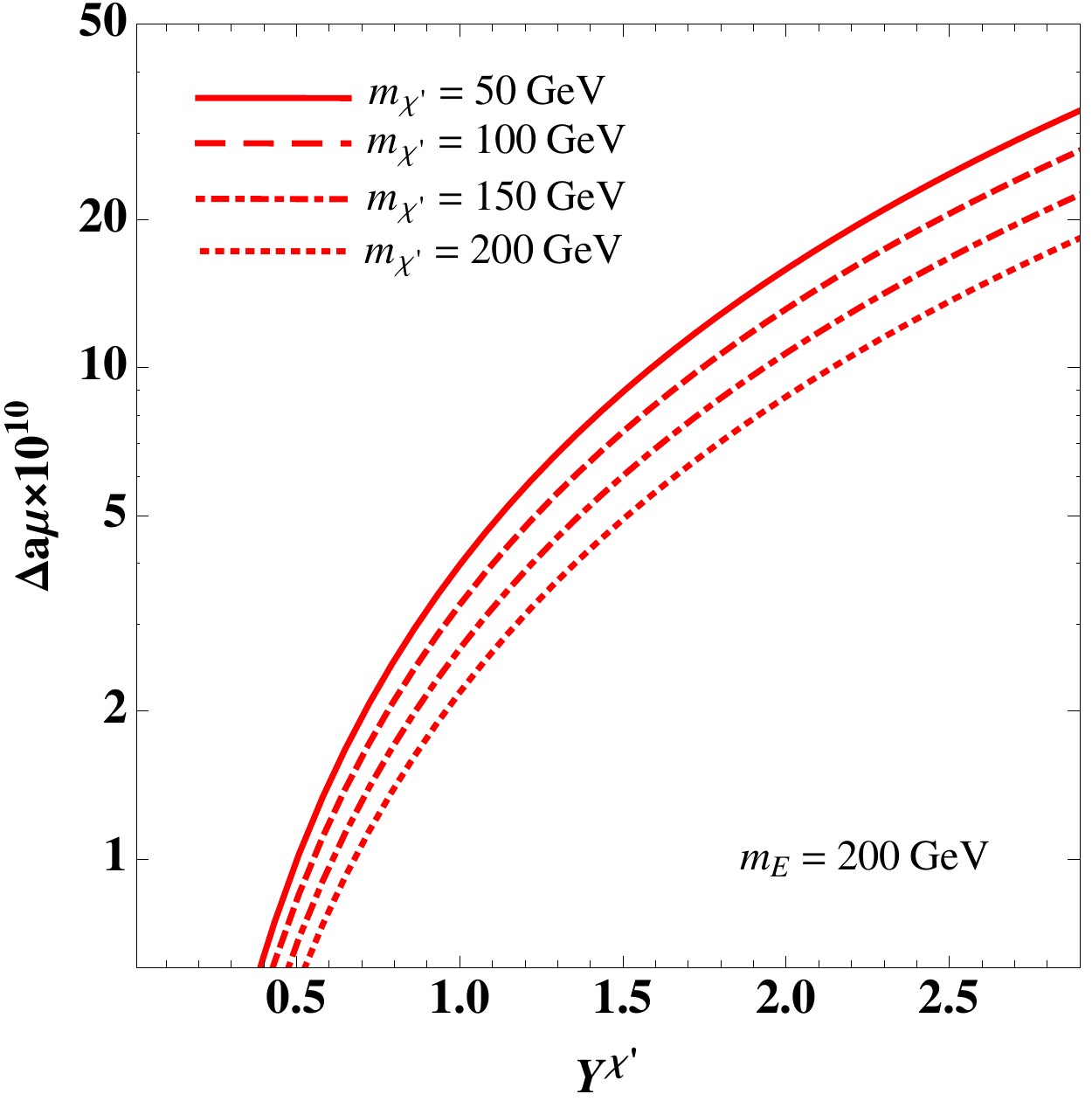}
\includegraphics[width=80mm]{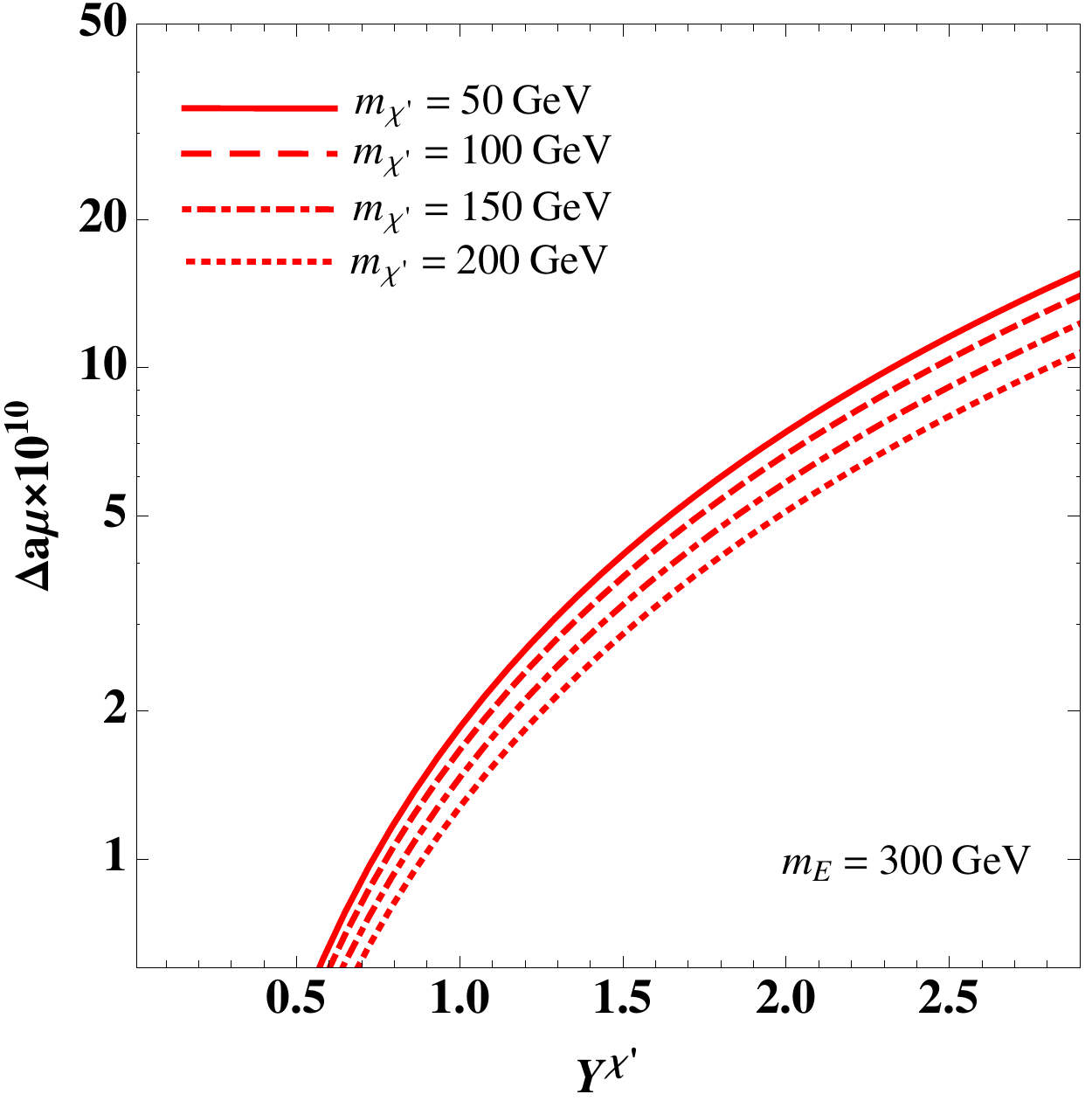} 
\caption{$\chi'$-$E$ loop contributions to $\Delta a_\mu$ as a function of Yukawa coupling $Y^{\chi'}$ where the masses of $\chi'$ and $E$ are indicated in the plots. }
\label{fig:Mg2b}
\end{figure}

In fact, we have a contribution to muon $g-2$ from one loop diagram in which $\chi'$ and $E$ propagate inside loop~\cite{Nomura:2019btk,Baek:2016kud}.
This contribution is estimated as 
\begin{align}
& \Delta a_\mu^{\rm 1 loop (\chi)} = \sum_{i=1-3} \frac{(Y^{\chi'})^T_{2i} Y^{\chi'}_{i 2}}{64 \pi^2} \frac{m_\mu^2}{M_{E_i}^2} F_{\chi'}(r_{E_i}), \\
& F_{\chi'} (r_{E_i}) = \int_0^1 d x \frac{x^2 (1-x)}{r_{E_i}^\mu x^2  + (1 - r_{E_i}^\mu) x + r_{E_i}^{\chi'} (1-x)}, 
\end{align}
where $r_{E_i}^\mu = m_\mu^2/M^2_{E_i}$ and $r_{E_i}^{\chi'} = m^2_{\chi'}/M^2_{E_i}$.
In Fig.~\ref{fig:Mg2b}, we show $\Delta a_\mu$ from $\chi'$-$E$ loop contribution as a function of Yukawa coupling $Y^{\chi'}$ where we assumed three generations of $E$ have the same mass
and all $Y^{\chi'}_{i2}$ has the same value. We also assume $Y^{\chi'}_{i 1(3)} = 0$ to avoid constraints from lepton flavor violation processes.
Thus $\Delta a_\mu \gtrsim 10^{-9}$ can be realized with sizable Yukawa coupling $Y^{\chi'}$ when the masses of $\chi'$ and $E$ are around electroweak scale.

\subsection{Collider physics}

In this subsection, we discuss collider physics mainly focusing of $Z'$ boson production at the LHC.
Our $Z'$ boson can be produced by $\bar q q \to Z'$ process since right-handed quarks have $U(1)_X$ charge.
We estimate the cross section using {\tt MADGRAPH/MADEVENT\,5}~\cite{Alwall:2014hca}, where the Feynman rules and relevant parameters in the model are implemented with FeynRules 2.0 \cite{Alloul:2013bka} and the {\tt NNPDF23LO1} PDF~\cite{Deans:2013mha} is adopted.
In the model $Z'$ can decay into SM quarks, scalar bosons and exotic charged lepton $E$ where branching ratio(BR) for $Z' \to  q \bar q$ is relatively larger than the other mode due to color degrees of freedom.
Then the most stringent constraint comes from the LHC analysis searching for $t \bar t$ resonance when $Z' \to t \bar t$ mode is kinematically allowed. 
When $m_{Z'} < 2 m_t$ our $Z'$ decays into jets and the collider constraint is looser due to large SM background cross section.
We can search such a $Z'$ boson by analyzing $pp \to \gamma Z'(\to jj/b\bar b)$ process with smaller number of SM backgrounds events.~\cite{Chiang:2015ika, Chiang:2014yva}

 \begin{figure}[tb]
\includegraphics[width=0.425\textwidth]{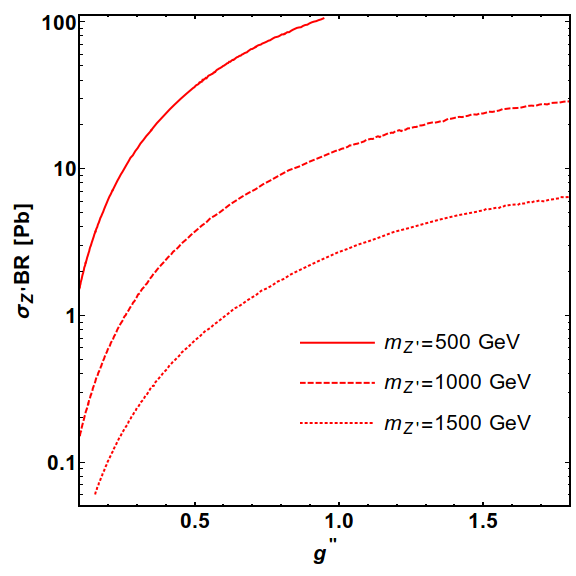}
\includegraphics[width=0.45\textwidth]{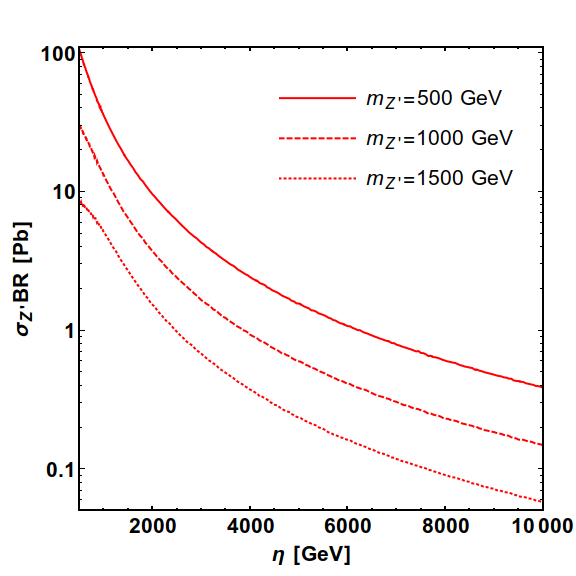}
\caption{Left: $Z'$ The products of production cross section and $BR(Z' \to t \bar t)$ as a function of $g''$ for $m_{Z'} = 500, 1000$ and $1500$ GeV. Right: $\sigma BR(Z' \to t \bar t)$ as a function of $\eta$ for the same values of $m_{Z'}$. }
\label{fig:CXZp}
\end{figure}

In the left(right) plot of Fig.~\ref{fig:CXZp} we show $\sigma(pp \to Z') BR(Z' \to  t \bar t)$ as a function of $g'' (\eta)$ for $m_{Z'} = \{500, 1000, 1500 \}$GeV. 
The estimated values of $\sigma(pp \to Z') BR(Z' \to  t \bar t)$ are compared with the upper bound from the analysis of LHC data~\cite{Aaboud:2018mjh} to search for allowed parameter region.
We then obtain allowed parameter space on $(M_{Z'}, \eta)$ plane where we also scanned $\tan \beta$ whose values are indicated by color gradient.
It is found that large $\eta$ region is allowed since $U(1)_X$ gauge coupling is small due to the relation $g'' \simeq M_{Z'}/\eta$. 
Furthermore $\tan \beta$ dependence is small since $Z$-$Z'$ mixing is always very small in the parameter region. 
In addition we estimate forward backward asymmetry (AFB) for $t \bar t$ final state from $Z'$ decay which is defined by
\begin{equation}
\Delta A_{FB} = \frac{N(\Delta |y| >0 ) -N(\Delta |y| < 0 ) }{N(\Delta |y| >0 ) + N(\Delta |y| < 0 )}
\end{equation}
where $N (\Delta |y| >(<) 0)$ indicates number of events with corresponding sign of $\Delta |y| = |y_t| - |y_{\bar t}|$ for rapidities of top and anti-top quarks $y_t$ and $y_{\bar t}$.
We find that $\Delta A_{FB} \sim 0.3-0.4$ is obtained in our model depending slightly on $Z'$ mass and it does not depend on the other parameters in the model.

 \begin{figure}[tb]
\includegraphics[width=0.8\textwidth]{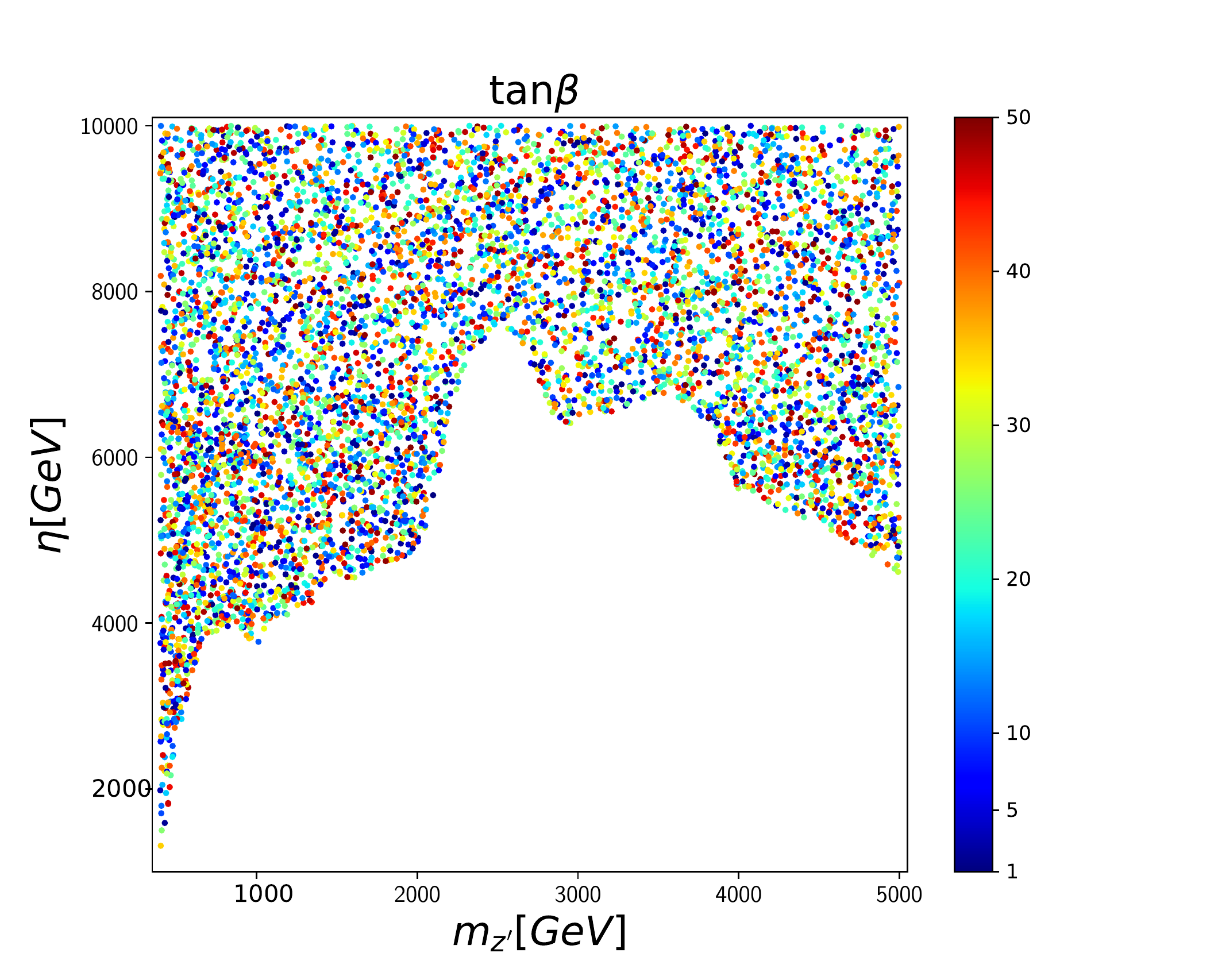}
\caption{The parameter region allowed by the constraint from data of $t \bar t$ search at the LHC where color gradient indicates the value of $\tan \beta$.}
\label{fig:Zp}
\end{figure}

% \begin{figure}[tb]
%\includegraphics[width=0.6\textwidth]{AFB_eta_15000_tb_1.pdf}
%\caption{$\Delta A_{FB}$ as a function of $Z'$ mass.}
%\label{fig:AFB}
%\end{figure}

\subsection{Dark matter physics}

Here we analyze DM physics such as relic density and constraint from direct/indirect detection experiments.
 In our model, DM candidates are $Z_2$ odd scalar bosons $\chi(\chi')$ and its interactions relevant to annihilation processes are given by
\begin{align}
\mathcal{L}  \supset \ & - i g'' Z'_\mu (\chi^* \partial^\mu \chi - \partial^\mu \chi^* \chi) + g''^2 Z'_\mu Z'^\mu \chi^* \chi  \nonumber \\
& - ( Y^{\chi'} \chi' \bar E_L e_R + h.c.)   - \frac{ \mu_\chi}{\sqrt{2}} [\chi \chi'(\phi_R+ i \phi_I) + h.c. ]  \nonumber \\
& \ + (\lambda_{\chi H_1} v \cos \beta h_1 + \lambda_{\chi H_2} v \sin \beta h_2 + \lambda_{\chi \phi} \eta \phi_R) \chi^* \chi \nonumber \\
& \ + (\lambda_{\chi' H_1} v \cos \beta h_1 + \lambda_{\chi' H_2} v \sin \beta h_2 + \lambda_{\chi' \phi} \eta \phi_R) \chi'^* \chi' \nonumber \\
&  + \frac{1}{2} \lambda_{\chi H_1} (h_1^2 + a_1^2 + \phi_1^+ \phi_1^-) \chi^* \chi + \frac{1}{2} \lambda_{\chi H_2} (h_2^2 + a_2^2 + \phi_2^+ \phi_2^-) \chi^* \chi
+  \frac{1}{2} \lambda_{\chi \phi} (\phi_R^2 + \phi_I^2 )\chi^* \chi \nonumber \\
&  + \frac{1}{2} \lambda_{\chi' H_1} (h_1^2 + a_1^2 + \phi_1^+ \phi_1^-) \chi'^* \chi' + \frac{1}{2} \lambda_{\chi' H_2} (h_2^2 + a_2^2 + \phi_2^+ \phi_2^-) \chi'^* \chi'
+  \frac{1}{2} \lambda_{\chi' \phi} (\phi_R^2 + \phi_I^2 )\chi'^* \chi',
\label{Eq:DMint}
\end{align}
where we ignored $Z$-$Z'$ mixing effect since it is negligibly small, and mass eigenstates for scalar fields are obtained applying Eqs.~(\ref{Eq:chargedMES}), (\ref{eq:CP-Odd}) and (\ref{Eq:CP-even}).
The scalar bosons and $Z'$ decay into SM particles via interactions given in Secs.~\ref{sec:scalar} and \ref{sec:Zp}.
Note that $\chi (\chi')$ decays into $\chi' \phi_{R,I}^{(*)} (\chi \phi_{R,I}^{(*)})$ state for $m_{\chi(\chi')} > m_{\chi'(\chi)}$ via the interaction  
with coupling $\mu_\chi$ so that only the lighter state among $\chi$ and $\chi'$ is the DM. 
Then we estimate relic density of our DM for each scenario given below applying {\tt micrOMEGAs 4.3.5}~\cite{Belanger:2014vza} by implementing relevant interactions.

In general we have many DM annihilation processes which are described by different parameters in Eq.~(\ref{Eq:DMint}).
Thus, in our analysis, we consider several scenarios focusing on some specific processes as follows: \\ 
 (1) $m_{\chi} < m_{\chi'}$ and  $\{ \lambda_{\chi \phi (\chi' \phi)},  \lambda_{\chi H_1 (\chi' H_1)}, \lambda_{\chi H_2 (\chi' H_2)}, Y^{\chi'} \} \ll 1$ so that $\chi \chi \to Z' \to f \bar f$ and/or $\chi \chi \to Z' Z'$ are dominant annihilation mode. \\
(2) $m_{\chi} < m_{\chi'}$ and $\{ \lambda_{\chi \phi},  \lambda_{\chi H_1}, \lambda_{\chi H_2} \}$ are sizable 
but $\{\lambda_{\chi' \phi},  \lambda_{\chi' H_1}, \lambda_{\chi' H_2}, Y^{\chi'}\} \ll 1 $ where scalar portal processes are dominant. \\
(3) $m_{\chi} > m_{\chi'}$ and $\{ \lambda_{\chi \phi (\chi' \phi)},  \lambda_{\chi H_1 (\chi' H_1)}, \lambda_{\chi H_2 (\chi' H_2)} \} \ll 1$ but $Y^{\chi'}$ is sizable where we consider $\chi' \chi' \to \ell^+ \ell^-$ process via Yukawa interaction. \\
Note that in scenario (2) we will get the same behavior if we exchange role of $\chi$ and $\chi'$ so that we only consider the case in which $\chi$ is DM. 
Under these scenarios, we estimate the relic density of DM. 

In addition to the relic density, we need to take into account constraints from DM direct detection experiments.
In our model DM can interact with nucleon through scalar and $Z'$ exchange when DM is $\chi$.
Then we can estimate the DM-nucleon scattering cross section, in non-relativistic limit, such that 
\begin{equation}
\sigma_{N-\chi} \simeq \frac{\mu_{N\chi}^2 m_N^2 f_N^2}{\pi s_\beta^2 m_\chi^2 v^2} 
\left( \frac{R_{22} C_{h^0 \chi \chi}}{m_h^2} + \frac{R_{21} C_{H^0 \chi \chi}}{m_{H^0}^2} + \frac{R_{23} C_{\xi^0 \chi \chi}}{m_{\xi^0}^2} \right)^2 + \frac{g''^4}{\pi} \frac{\mu_{N \chi}^2}{m_{Z'}^4}, 
\end{equation}
where $m_N$ is nucleon mass, $\mu_{N\chi} = m_N m_\chi/(m_N + m_\chi)$ and $f_N$ is effective Nucleon-Higgs coupling~\cite{He:2011de, Cheng:2012qr}.
The couplings $C_{[H^0,h^0,\xi^0] \chi \chi}$ are obtained from terms in second line of Eq.~(\ref{Eq:DMint}) such that
\begin{align}
& C_{[H^0, h^0, \xi^0 ] \chi \chi} = \lambda_{\chi H_1} v \cos \beta R_{1[1,2,3]} + \lambda_{\chi H_2} v \sin \beta R_{2[1,2,3]} + \lambda_{\chi \phi} \eta R_{3[1,2,3]}.
\label{Eq:coupling_DM_scalar}
\end{align}
In our numerical analysis below, we adopt {\tt micrOMEGAs 4.3.5} in estimating $\sigma_{N-\chi}$ and the experimental constraints are imposed~\cite{Aprile:2018dbl}.
When DM is $\chi'$ only scalar mediating interaction contribute to DM-nucleon scattering where we can obtain the contribution by exchanging $\chi$ to $\chi'$ for couplings in Eq.~(\ref{Eq:coupling_DM_scalar}).

\begin{figure}[tb]
\includegraphics[width=0.6\textwidth]{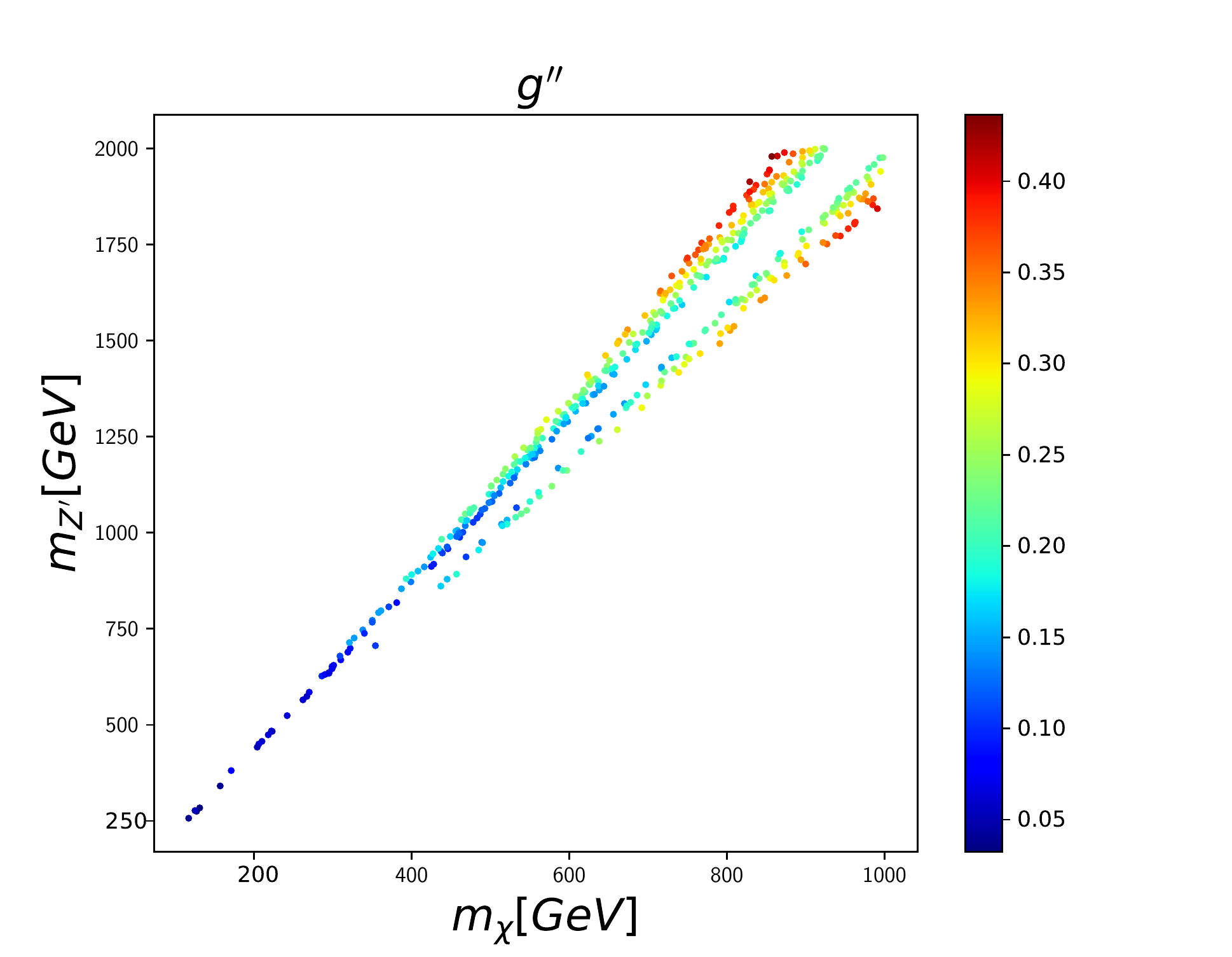}
\caption{Parameter region accommodating with relic density of DM in scenario (1)}
\label{fig:DM_allowed_case1}
\end{figure}

\begin{figure}[tb]
\includegraphics[width=0.6\textwidth]{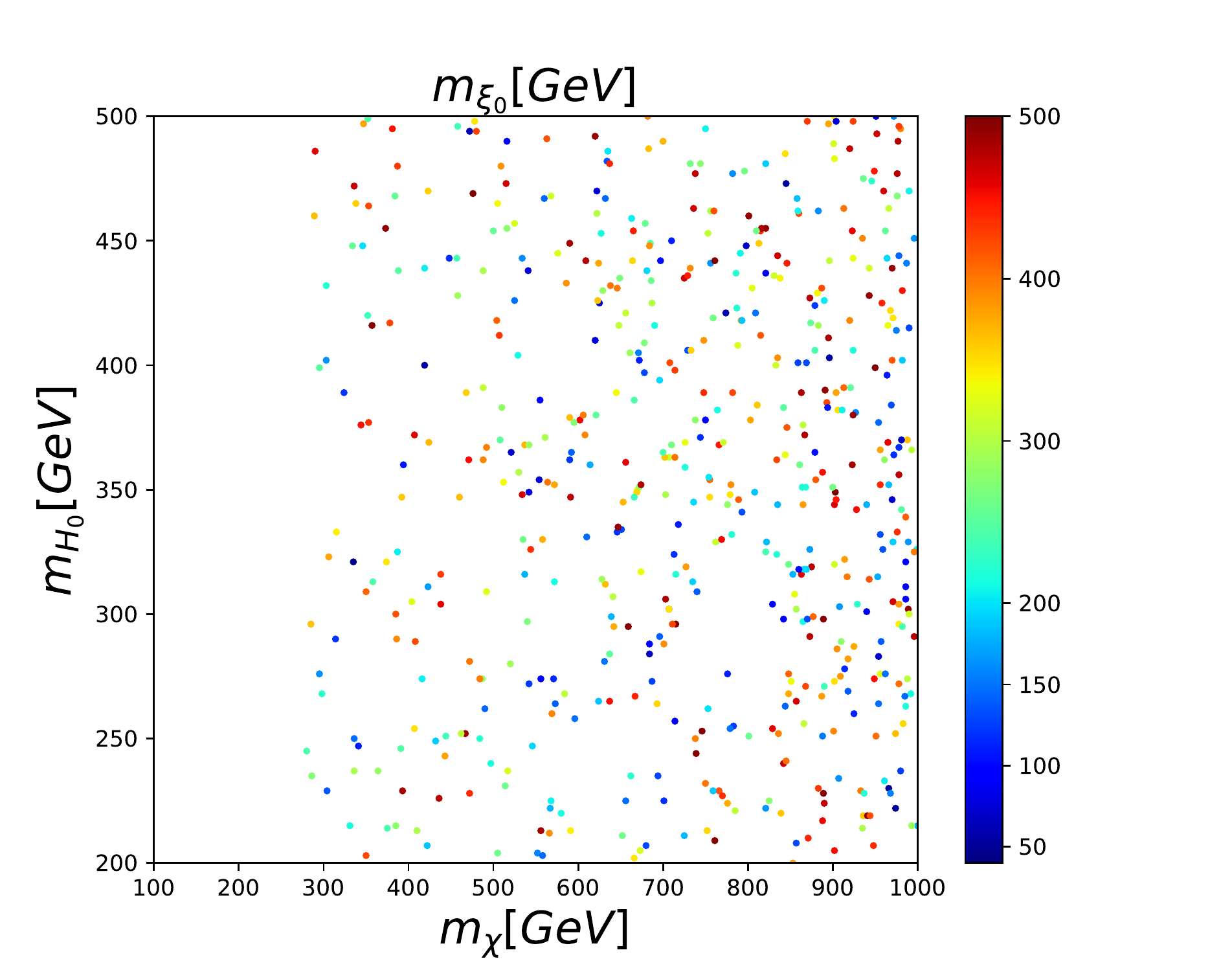}
\caption{Parameter region accommodating with relic density of DM in scenario (2).}
\label{fig:DM_allowed_case2}
\end{figure}

 \begin{figure}[tb]
\includegraphics[width=0.6\textwidth]{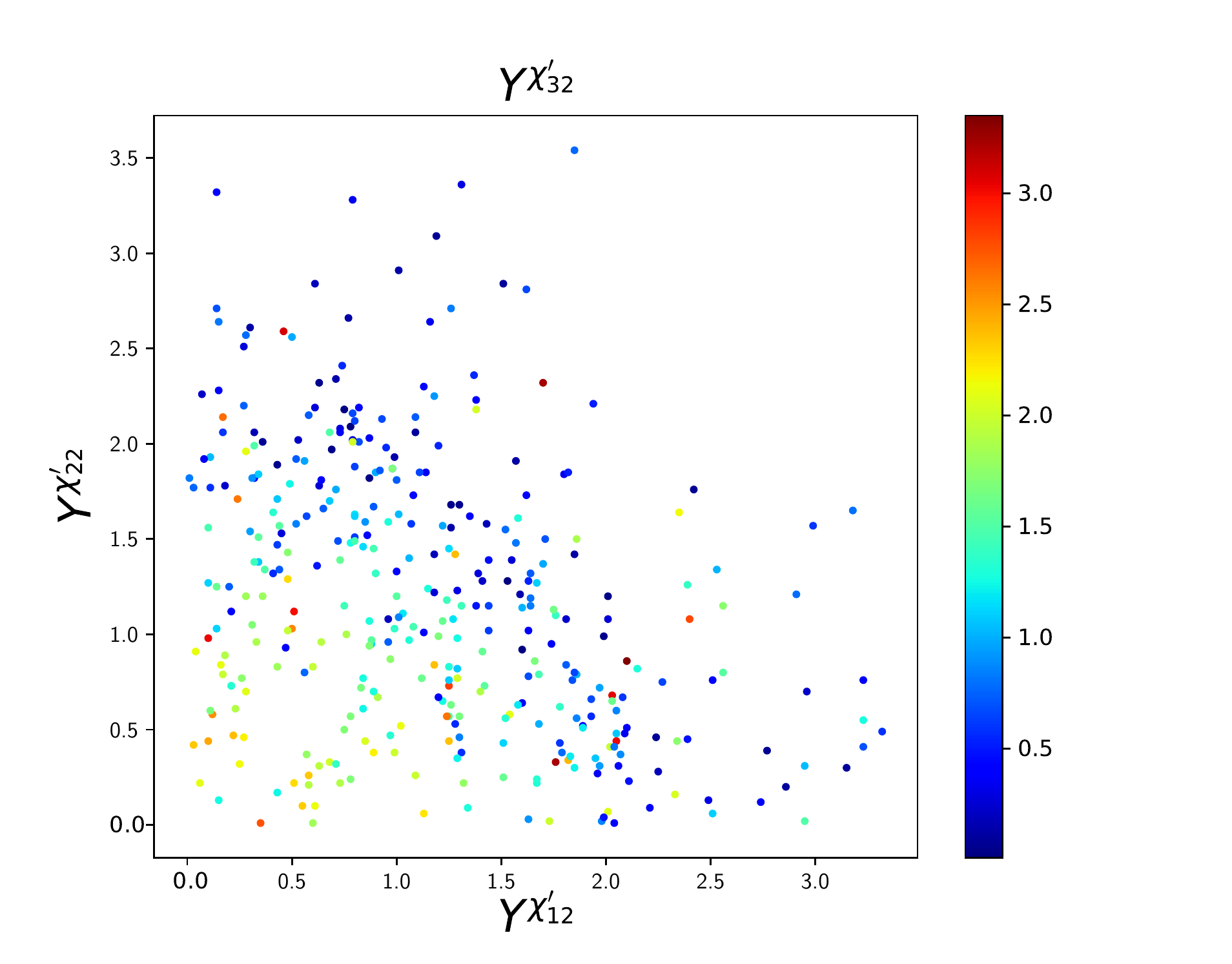}
\caption{Parameter region accommodating with relic density of DM in scenario (3).}
\label{fig:DM_allowed_case3}
\end{figure}

We perform parameter scan for each scenarios to search for parameter region realizing observed relic density of DM.
Firstly we set following parameter ranges for all scenarios:
\begin{align}
& m_{\chi (\chi')} \in [100, 1000] \ {\rm [GeV]}, \quad m_{H^0} = m_{H^\pm} \in [200, 500] \ {\rm [GeV]}, \nonumber \\
& m_{A^0, \xi^0} \in [40, 500] \ {\rm [GeV]}, \quad M_{E_i} \in [m_\chi, 1200] \ {\rm [GeV]} \nonumber \\
& \tan \beta \in [1, 50], \quad \alpha_1 \in \left[- \frac{\pi}{2}, 0 \right], \quad \alpha_2 \in \left[ -\frac{\pi}{2}, \frac{\pi}{2} \right], \quad \alpha_3 \in - {\rm sign}(\alpha_2) \times \left[0, \frac{\pi}{2} \right], 
\end{align}
where the range of $\alpha_i$ is chosen as indicated by the constraints from scalar sector discussed above.
The other parameters are set for scenario (1) as
\begin{align}
& M_{Z'} \in [150, 2000] \ {\rm [GeV]}, \quad \eta \in [4500, 10000] \ {\rm [GeV]}, \nonumber \\
&  \lambda_{\chi \phi (\chi' \phi)} = \lambda_{\chi H_1 (\chi' H_1)} = \lambda_{\chi H_2 (\chi' H_2)} = 10^{-5}, \quad Y_{ij}^{\chi'} = 10^{-5}.
\end{align}
In addition, we impose LHC constraint on $\{m_{Z'}, g'' \}$ parameter space discussed in the previous section and we scan these values within allowed region. 
For scenario (2), we chose 
\begin{align}
& M_{Z'} = 2500 \ {\rm [GeV]}, \quad \eta = 10000 \ {\rm [GeV]}, \nonumber \\
& \{ \lambda_{\chi \phi}, \lambda_{\chi H_1}, \lambda_{\chi H_2} \} \in [0.001, 0.1], \quad \lambda_{\chi' \phi} = \lambda_{\chi' H_1} = \lambda_{\chi' H_2} = 10^{-5}, \quad Y_{ij}^{\chi'} = 10^{-5}.
\end{align}
For scenario (3), we chose 
\begin{align}
& M_{Z'} = 2500 \ {\rm [GeV]}, \quad \eta = 10000 \ {\rm [GeV]}, \nonumber \\
&  \lambda_{\chi \phi (\chi' \phi)} = \lambda_{\chi H_1 (\chi' H_1)} = \lambda_{\chi H_2 (\chi' H_2)} = 10^{-5}, \quad Y_{i2}^{\chi'} \in [0.01, \sqrt{4 \pi}],\quad Y_{ij}^{\chi'}=10^{-5} \hspace{1.5mm} \text{for} \hspace{1.5mm} j\neq 2.
\end{align}
Note that we assume $\chi$ and $\chi'$ masses are not degenerated, and co-annihilation processes are not taken into account in relic density calculation.

In Fig.~\ref{fig:DM_allowed_case1}, we show allowed parameter region giving relic density, $0.11 < \Omega h^2 < 0.13$, in scenario (1) where horizontal(vertical) axis corresponds to $M_{\chi}(M_{Z'})$ and color gradient indicate the value of $g''$. It is found that the observed relic density can be obtained around $M_{Z'} \sim 2 M_{\chi}$ since the annihilation cross section is enhanced by resonant effect~\cite{Griest:1990kh, Ibe:2008ye,Nayak:2017dwg,Pozzo:2018anw,Athron:2018ipf}.
We also show allowed parameter region for scenario (2) in Fig.~\ref{fig:DM_allowed_case2} where horizontal(vertical) axis indicates $M_{\chi} (M_{H_0})$ and color gradient shows $M_{\xi_0}$.
In this case we obtain allowed region for $M_{\chi} < M_{H_0}$(or $M_{\xi_0}$) since the relic density is explained by the process $\chi \chi \to H_0 H_0 (\xi_0 \xi_0)$.
In addition, the allowed parameter region for scenario (3) is shown in Fig.~\ref{fig:DM_allowed_case3} where horizontal(vertical) axis indicates $Y^{\chi'}_{12} (Y^{\chi'}_{22})$ and color gradient shows $Y^{\chi'}_{32}$.
We find that required values of Yukawa couplings $Y^{\chi'}_{i2}$ are $\mathcal{O}(1)$ scale which is also required to obtain sizable $\Delta a_\mu$.

%%%%%%%%%%%%%%%%%%%%%%%%%%%%%%%%%%%%%%%%%%%%%%%%%%%
\begin{figure}[tb]
\includegraphics[width=0.6\textwidth]{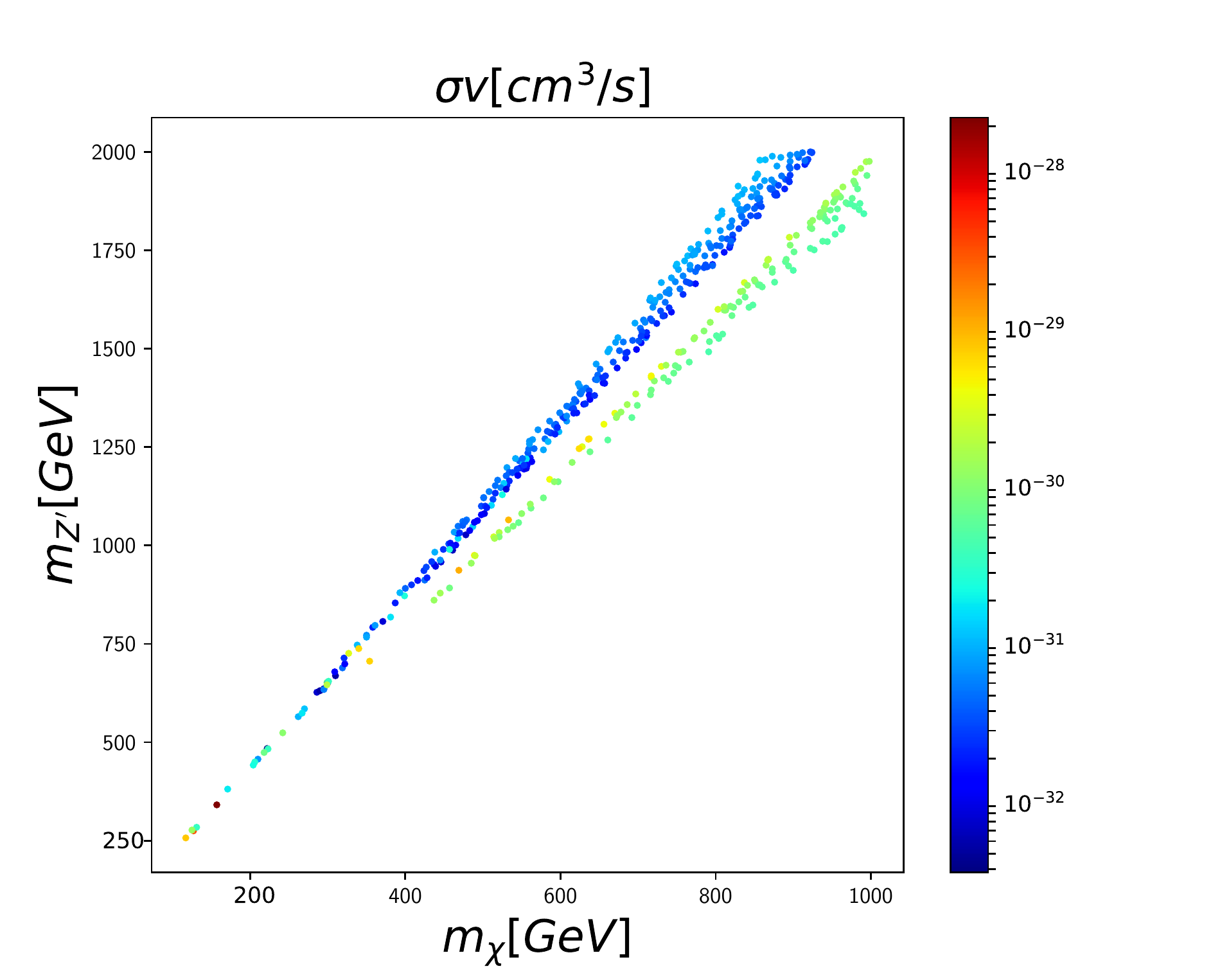}
\caption{DM annihilation cross section at the current universe in scenario (1)}
\label{fig:DM_sigmav_case1}
\end{figure}

\begin{figure}[tb]
\includegraphics[width=0.6\textwidth]{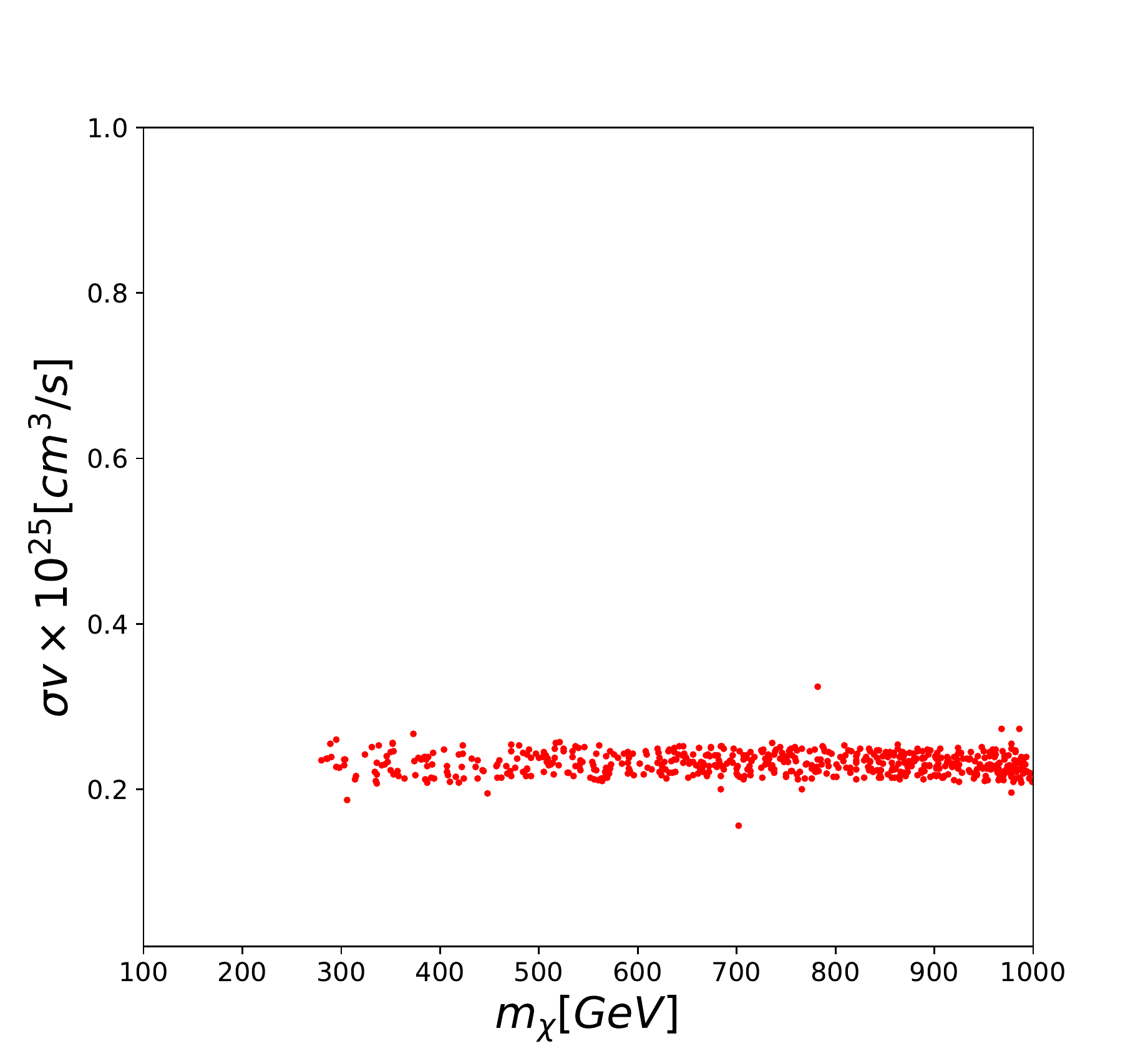}
\caption{DM annihilation cross section at the current universe in scenario (2).}
\label{fig:DM_sigmav_case2}
\end{figure}
%%%%%%%%%%%%%%%%%%%%%%%%%%%%%%%%%%%%%%%%%%%%%%%%%%%

\begin{table}[t]
  \begin{center}
    \begin{tabular}{|c|c|c|c|c} \hline 
      \hspace{2.5mm} Parameters \hspace{2.5mm} &\hspace{6mm} BPI \hspace{6mm} & \hspace{6mm} BPII \hspace{6mm} \\
      \hline
      $m_{\chi}$ & 590 GeV & 766 GeV\\
      $m_{H^0}$ & 362 GeV & 368 GeV\\
      $m_{H_\pm}$ & 362 GeV & 368 GeV\\
      $m_{A^0}$ & 283 GeV & 143 GeV \\
      $m_{\xi^0}$ & 121 GeV & 449 GeV \\
      $m_{E_1}$ & 971 GeV & 811 GeV \\
      $m_{E_2}$ & 673 GeV & 915 GeV \\
      $m_{E_3}$ & 954 GeV & 1143 GeV \\
      $\tan\beta$ & 2.8 & 13.8 \\
      $\alpha_1$ & $-1.51$ rad & $-0.39$ rad \\
      $\alpha_2$ & $-0.09$ rad & 1.34 rad \\
      $\alpha_3$ & 1.55 rad & $-0.34$ rad \\
      $\lambda_{\chi \phi}$ & 0.042 & 0.053 \\
      $\lambda_{\chi H_1}$ & 0.077 & 0.062 \\
      $\lambda_{\chi H_2}$ & 0.018 & 0.041 \\
      \hline
    \end{tabular}
  \end{center}
  \caption{Parameter choice allowed by relic density and direct detection constraints for computing the photon flux for three angular regions.}
      \label{tab:table3}
\end{table}
%%%%%%%%%%%%%%%%%%%%

%%%%%%%%%%%%%%%%%%%%%%%%%%%%%%%%%%%%%%%%%%%%%%%%%%%
\begin{figure}[tb]
\includegraphics[width=0.45\textwidth]{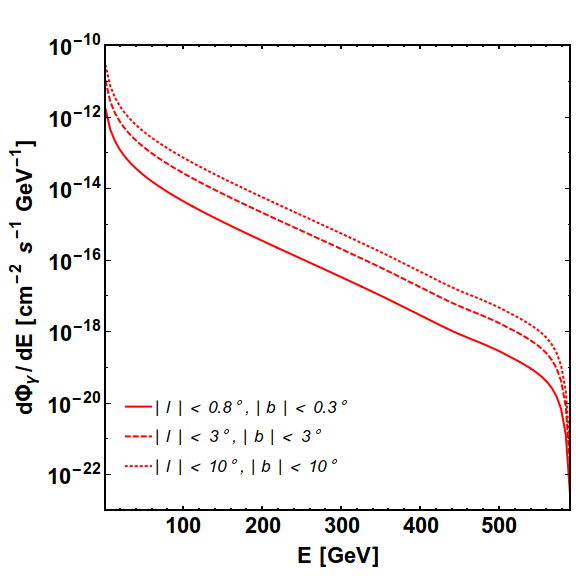}
\includegraphics[width=0.45\textwidth]{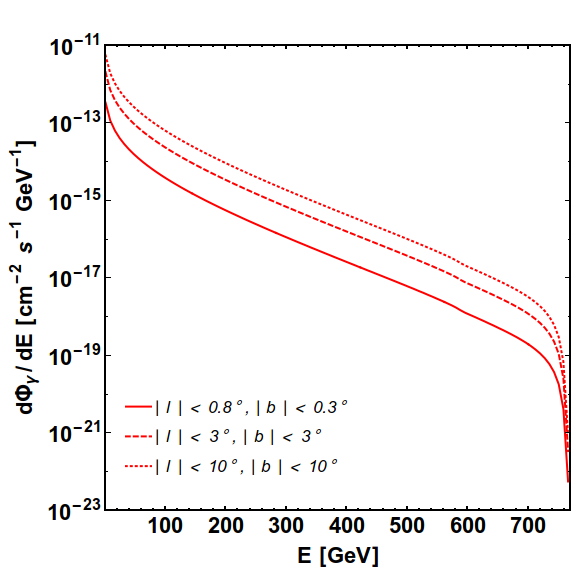}
\caption{Left and right plots correspond to $\gamma$-ray spectrum from DM annihilation for two benchmark points 1 and 2 in scenario (2) where we applied three angular regions.}
\label{fig:photon_spectrum}
\end{figure}
%%%%%%%%%%%%%%%%%%%%%%%%%%%%%%%%%%%%%%%%%%%%%%%%%%%%

Finally we comment on possibility of indirect detection of our DM. 
For each scenario above, DM pair annihilates mainly as follows: in scenario (1) $\chi \bar \chi \to Z' \to f_{SM}  \bar f_{SM}$ or $\chi \bar \chi \to Z' Z' \to 2 f_{SM}  \bar f_{SM}$;
in scenario (2) $\chi \bar \chi \to \phi^0 \to f_{SM}  \bar f_{SM}$ or $\chi \bar \chi \to \phi \phi \to 2 f_{SM}  \bar f_{SM}$ where $\phi^0$ and $\phi$ indicate neutral scalar and any scalar bosons;
in scenario (3) $\chi' \chi' \to \ell^+ \ell^-$ where $\ell$ is the SM lepton.
Then gamma-ray search gives the strongest constraint on the annihilation cross section by Fermi-LAT observation~\cite{Hoof:2018hyn, Fermi-LAT:2016uux}.
For scenario (3), current DM annihilation is small since the cross section is suppressed by DM velocity since it is P-wave dominant process.
We thus estimate DM annihilation cross section in current universe for scenario (1) and (2) using {\tt micrOMEGAs 4.3.5}.
In Fig.~\ref{fig:DM_sigmav_case1} and \ref{fig:DM_sigmav_case2}, we respectively show the DM annihilation cross section in the current universe for scenario (1) and (2).
We find that the cross section in scenario (1) is smaller than that in scenario (2) since DM-DM-$Z'$ coupling include derivative and the cross section is suppressed by momentum factor.
Therefore the scenario (2) is the most sensitive case for indirect detection where the shown parameter region is still allowed by the current measurements~\cite{Hoof:2018hyn, Fermi-LAT:2016uux},
 and it can be tested by in future data.   
 For illustration, we also estimate spectrum of $\gamma$-ray from DM annihilation in scenario (2) where we adopt two benchmark points(BPs) given in Table~\ref{tab:table3} and use {\it micrOMEGAs}. 
The spectrum for BP1 and BP2 are shown in left and right plot of Fig.~\ref{fig:photon_spectrum} where we applied three angular regions characterized by galactic latitude $b$ and longitude $l$.
We find that $d \Phi_\gamma/dE$ has broad range and its value is larger for smaller energy region 
since $\gamma$-ray comes from radiation from charged particle in final states in DM annihilation.

\section{Summary and Conclusions}

We have constructed a two Higgs doublet model with extra $U(1)_X$ gauge symmetry in which 
lepton specific (type-X) structure is realized by charge assignment of Higgs doublets, quarks and leptons.
In addition exotic charged leptons $E$ are also introduced to cancel gauge anomalies.
We have also introduced discrete $Z_2$ symmetry under which exotic charged leptons have odd parity, in order to restrict exotic charged lepton interactions.
Furthermore the SM singlet scalars $\chi$ and $\chi'$ with $Z_2$ odd parity are added as our dark matter candidate where $\chi$ is charged under $U(1)_X$ while $\chi'$ is not charged.

We analyzed scalar sector formulating mass eigenstates and relation among parameters in the scalar potential. 
Then allowed parameter regain is explored by investigating constraints from scalar sector such as stability and peturbativity bound in the potential.
We have also estimated muon $g-2$ applying the allowed parameter sets.
It has been found that the contributions from loop diagrams with $Z_2$ even scalar bosons can not be sizable to explain muon $g-2$ discrepancy.
To explain muon $g-2$ we should rely on contribution from loop diagrams with $Z_2$ odd particle and it can give sufficiently large muon $g-2$ with sizable Yukawa coupling associated with exotic leptons and dark matter.

The collider physics has been also discussed focusing on $Z'$ boson production at the LHC.
Our $Z'$ boson has leptophobic interactions and $pp \to Z' \to t \bar t$ process provides the strongest constraint if $Z'$ mass is heavier than $2 m_t$.
We have estimated the $Z'$ production cross section and discussed its constraints.
In addition we have discussed $t \bar t$ asymmetry for the $pp \to Z' \to t \bar t$ process.

Finally we have analyze dark matter physics such as relic density and constraint from direct/indirect detection experiments.
In our analysis, we have considered several scenarios: (1) DM is $\chi$ and $Z'$ interaction is dominant, (2) DM is $\chi$ (or $\chi'$) and scalar portal interaction is dominant, (3) DM is $\chi'$ and Yukawa interaction with exotic leptons is dominant.
Then allowed parameter region for each case have been searched for taking into account observed relic density and direct detection constraints.
We then find all the cases can realize the observed relic density by choosing parameters relevantly.
In addition we have discussed possibility of indirect detection estimating DM annihilation cross section at the current universe. 
It has been shown that scenario (2) is the most sensitive to indirect detection and will be tested in future measurements.

%%%%%%%%%%%%%%%%%%%%%%%%%%%%%%%%%%%
%\section*{Acknowledgments}
%\vspace{0.3cm}
%%%%%%%%%%%%%%%%%%%%%%%%%%%%%%%%%%%

\end{document}